\documentclass[amsmath,amssymb]{nature}

\usepackage{bm}
\usepackage{color}
\usepackage[normalem]{ulem}
\usepackage{newfloat}
\usepackage[latin1]{inputenc}
\usepackage{comment}
\usepackage{amsmath}
\usepackage[dvipdfmx]{hyperref}

\newcommand{\rev}[1]{\textcolor{blue}{#1}}

\usepackage[dvipdfmx]{graphicx}

\DeclareFloatingEnvironment[name={Supplementary Figure S}]{suppfigure}

\title{Evidence for Magnetic Weyl Fermions in a Correlated Metal} 

\author{K.~Kuroda$^1$\footnote[2]{These two authors contributed equally.}, T.~Tomita$^{1,3}$\footnotemark[2], M.-T.~Suzuki$^{2,3}$, C.~Bareille$^1$, A.~A.~Nugroho$^{1,4}$, P.~Goswami$^5$, M.~Ochi$^6$, M.~Ikhlas$^{1,3}$, M.~Nakayama$^1$, S.~Akebi$^1$, R.~Noguchi$^1$, R.~Ishii$^1$, N.~Inami$^7$, K.~Ono$^7$, H.~Kumigashira$^7$, A. Varykhalov$^8$, T.~Muro$^9$, T.~Koretsune$^{2,3}$, R.~Arita$^{2,3}$, S.~Shin$^1$, Takeshi~Kondo$^1$, S.~Nakatsuji$^{1,3,*}$}

\begin{document}

\maketitle

\begin{affiliations}
\item Institute for Solid State Physics, University of Tokyo, Kashiwa 277-8581, Japan
\item RIKEN Center for Emergent Matter Science (CEMS), 2-1 Hirosawa, Wako, Saitama 351-0198, Japan
\item CREST, Japan Science and Technology Agency (JST), 4-1-8 Honcho Kawaguchi, Saitama 332-0012, Japan
\item Faculty of Mathematics and Natural Sciences, Institut Teknologi Bandung, Jl. Ganesha 10, 40132 Bandung, Indonesia
\item Condensed Matter Theory Center and Joint Quantum Institute, Department of Physics, University of Maryland, College Park, Maryland 20742- 4111 USA
\item Department of Physics, Osaka University, Machikaneyama-cho, Toyonaka, Osaka 560-0043, Japan
\item Institute of Materials Structure Science, High Energy Accelerator Research Organization (KEK), Tsukuba, Ibaraki 305-0801, Japan
\item Helmholtz-Zentrum Berlin f{\"u}r Materialien und Energie, Elektronenspeicherring BESSY II, Albert-Einstein-Strasse 15, 12489 Berlin, Germany
\item JASRI/SPring-8, Kouto 1-1-1, Sayo-cho, Sayo-gun, Hyogo 679-5198, Japan
\end{affiliations}

\begin{abstract}
Recent discovery of both gapped and gapless topological phases in weakly correlated electron systems has introduced various relativistic particles and a number of exotic phenomena in condensed matter physics~\cite{hasan10rmp,Liu14science,Neupane14,Xu15science,Lv2015}. The Weyl fermion\cite{Nielsen83plb,Wan11prb, Burkov2011} is a prominent example of three dimensional (3D), gapless topological excitation, which has been experimentally identified in inversion symmetry breaking semimetals~\cite{Xu15science,Lv2015}. However, their realization in spontaneously time reversal symmetry (TRS) breaking magnetically ordered states of correlated materials has so far remained hypothetical~\cite{Wan11prb,Witczak2014,Wang16prl}.
Here, we report a set of experimental evidence for elusive magnetic Weyl fermions in Mn$_3$Sn, a non-collinear antiferromagnet that exhibits a large anomalous Hall effect even at room temperature~\cite{Nakatsuji15Nature}. Detailed comparison between our angle resolved photoemission spectroscopy (ARPES) measurements and density functional theory (DFT) calculations reveals significant bandwidth renormalization and damping effects due to the strong correlation among Mn 3$d$ electrons. Moreover, our transport measurements have unveiled strong evidence for the chiral anomaly of Weyl fermions, namely, the emergence of positive magnetoconductance only in the presence of parallel electric and magnetic fields.
The magnetic Weyl fermions of Mn$_3$Sn have a significant technological potential, since a weak field ($\sim$ 10~mT) is adequate for controlling the distribution of Weyl points and the large fictitious field ($\sim$ a few 100~T) in the momentum space. Our discovery thus lays the foundation for a new field of science and technology involving the magnetic Weyl excitations of strongly correlated electron systems.
\end{abstract}
%
Traditionally, topological properties have been considered for the systems supporting gapped bulk excitations~\cite{hasan10rmp}. However, over the past few years three dimensional gapless systems such as Weyl and Dirac semimetals have been discovered, which combine two seemingly disjoint notions of gapless bulk excitations and band topology~\cite{Liu14science,Neupane14,Xu15science,Lv2015}. In 3D inversion or TRS breaking systems, two nondegenerate energy bands can linearly touch at pairs of isolated points in the momentum ($\mathbf{k}$) space, giving rise to the Weyl quasiparticles. The touching points or Weyl nodes act as the unit strength (anti) monopoles of underlying Berry curvature\cite{Xu15science,Lv2015,Nielsen83plb,Wan11prb}, leading to the protected zero energy surface states also known as the Fermi-arcs~\cite{Wan11prb,Xu15science,Lv2015}, and  many exotic bulk properties such as large anomalous Hall effect (AHE)\cite{Yang2011}, optical gyrotropy\cite{Shudan2015}, and chiral anomaly\cite{Nielsen83plb,Son2013,Goswami2015,Xiong2015,Huang2015,Zhang2016,Hirschberger16naturemat}. Interestingly, the Weyl fermions can describe low energy excitations of both weakly and strongly correlated electron systems. In weakly correlated, inversion symmetry breaking materials, where the symmetry breaking is entirely caused by the crystal structure rather than the collective properties of electrons, the ARPES has provided evidence for long-lived bulk Weyl fermions and the surface Fermi arcs~\cite{Xu15science,Lv2015}. On the other hand, the magnetic Weyl fermions have been predicted for several spontaneously TRS breaking phases of strongly correlated materials since the early stage of the search~\cite{Wan11prb,Witczak2014,Wang16prl}, but so far they have evaded experimental detection. In correlated electron systems, when the spectroscopic detection of coherent Weyl quasiparticles and Fermi arcs can be complicated due to the interaction induced suppression of bandwidth and life time, it becomes essential to perform complementary measurements of bulk physical quantities such as AHE and chiral anomaly, which are sensitive to the underlying topology. 
 
Among the candidates, Mn$_3$Sn is the only compound that exhibits a spontaneous Hall effect~\cite{Nakatsuji15Nature}. Mn$_3$Sn is a hexagonal antiferromagnet (AFM) with a stacked kagome lattice. The geometrical frustration leads to a 120$^{\circ}$ structure of Mn moments, whose symmetry allows a very small spin canting and thus macroscopic magnetization (Fig.~\ref{fig:fig1}a)~\cite{Tomiyoshi1982,Brown90jpcm}. This is the first AFM that exhibits a surprisingly large AHE below the N{\'{e}}el temperature of 430~K, which is comparable with or even exceeds the AHE of ferromagnets even though the magnetization is negligibly small. Such a large AHE in an AFM at room temperature is not only potentially relevant for technological applications, but indicates a novel mechanism~\cite{Suzuki17prb} that induces a large Berry curvature $\mathbf{\Omega(k)}$ \cite{Nagaosa10rmp}, whose scale reaches $\sim$ a few 100 T ~\cite{Kiyohara2016}.
In fact, a recent band calculation has shown that the antiferromagnetic state can support Weyl fermions and large $\mathbf{\Omega(k)}$ ~\cite{Yang2017njp}.
Here, we show that Mn$_3$Sn is strongly correlated by ARPES measurements. Moreover, we provide evidence for magnetic Weyl fermions by combining the observations of large AHE\cite{Nakatsuji15Nature}, and concomitant positive longitudinal magnetoconductance (LMC) and negative transverse magnetoconductance (TMC) originating due to the chiral anomaly.

%
%
We first present in Fig.~\ref{fig:fig1}b the overview of the band structure calculated with spin-orbit coupling (SOC) for the magnetic configuration shown in Fig.\ref{fig:fig1}a. 
The TRS breaking lifts the spin degeneracy and leads to the band crossing at a number of $\mathbf{k}$ points, resulting in various pairs of the Weyl nodes at different energies.
Among them, the most relevant for transport and other macroscopic measurements are the Weyl points that are closest to the Fermi energy, $E_{\rm{F}}$.
In accordance with the previous prediction~\cite{Yang2017njp}, such Weyl points are found along ${\rm{K}}$-${\rm{M}}$-${\rm{K}}$ line (dotted rectangle in Fig.~\ref{fig:fig1}b); an electron band and a hole band centered at ${\rm{M}}$ intersect slightly above $E_{\rm{F}}$ forming the Weyl points.
Moreover, we show in the following that the magnetic texture sensitively determines the presence/absence of the gap at the band crossing.

In Fig.~\ref{fig:fig1}c, we summarize the $k_{x}$-$k_{y}$ location of the above mentioned Weyl nodes. 
Without SOC, the electron-hole band crossings form a nodal ring surrounding $\rm{K}$ points (dashed circles).
Inclusion of SOC opens the gap along the nodal ring, except a pair of points corresponding to the Weyl nodes with different chirality. Notably, in Mn$_3$Sn, the direction of the sublattice moments can be controlled by a small magnetic field of $\sim$ a few 100~Oe (Supplementary Note~1 and Figure~S1).
Thus, by rotating a magnetic field in the $x$-$y$ plane, the Weyl nodes, whose locations are determined by the spin texture and thus by magnetic field, may be moved along the hypothetical nodal ring.

As an important consequence of the magnetic symmetry, the electronic structure becomes orthorhombic even in the hexagonal crystal system.
The presence of the mirror symmetry ensures that the pairs of Weyl nodes appear
along a $\mathbf{k}$ line parallel to the local easy-axis (i.e. $x$ axis).
To demonstrate this, we show in Fig.~\ref{fig:fig1}d an enlarged view of the band structure cut along the two distinct lines, $\rm{K}$-$\rm{M}$-$\rm{K}$ and $\rm{K}$-$\rm{M'}$-$\rm{K}$ (Fig.~\ref{fig:fig1}c).
For $\rm{K}$-$\rm{M}$-$\rm{K}$, the electron-hole band crossing generates the Weyl nodes (W$_1^{+}$, W$_2^{-}$) at $E-E_{\rm{F}} \sim 60$~meV, which is however absent for $\rm{K}$-$\rm{M'}$-$\rm{K}$ due to the interband SOC coupling.
Interestingly, their counterparts (W$_2^{+}$, W$_1^{-}$) arise slightly above this energy, at $\sim 90$~meV.
Since the Weyl nodes appear at the points where electron and hole pockets intersect, they give rise to a type-II Weyl semimetal~\cite{Soluyanov2015nature}.
The presence of the electron and hole bands around $\rm{M}$ is thus a key for realizing magnetic Weyl fermions in Mn$_{3}$Sn.
Recent extensive studies have shown that, in weakly correlated systems, the DFT calculation can excellently reproduce non-trivial band topology found by ARPES measurements~\cite{hasan10rmp,Liu14science,Neupane14,Xu15science,Lv2015}. 
To directly verify the 3D electronic structure in Mn$_{3}$Sn, we performed ARPES measurements with tunable photon energy using synchrotron radiation (Fig.~\ref{fig:fig2}). The single crystals used here are slightly off stoichiometric with extra Mn (Mn$_{3.03}$Sn$_{0.97}$), which should move up $E_{\rm F}$ by 19~meV according to theory (Figs.~\ref{fig:fig1}b \& d, Methods). Hereafter, we focus on the coplanar magnetic phase at $T >$ 50 K\cite{Nakatsuji15Nature}. 
By systematically changing the photon energy ($h\nu$) from 50 to 170~eV, we observe the clear $k_{z}$ dispersion along $\rm{H}$-$\rm{K}$-$\rm{H}$ line with the quasiparticle peak developed near $E_{\rm{F}}$ around $\rm{K}$ (blue arrow in Fig.~\ref{fig:fig2}c).
This clarifies that the incident $h\nu$ of 103~eV selectively detects the bulk band dispersion at $k_{z}=0$, where the Weyl nodes should exist near $E_{\rm{F}}$.
The contour of photoelectron intensity clarifies the location of the Fermi surfaces on the $k_{x}$-$k_{y}$ plane at $k_{z}=0$ (Fig.~\ref{fig:fig2}b).
Experiment clearly captures the main six elliptical-shaped contours centered at $\rm{M}$ points, which have the same topology as the Fermi surfaces (solid circle) predicted by DFT. 
This agreement is significant as it is this electron band that creates the Weyl points at its intersection with the other hole band (Fig.~\ref{fig:fig1}d).

The ARPES intensity maps and their energy distribution curves 
show the band dispersion along $\rm{K}$-$\rm{\Gamma}$-$\rm{K}$ (Fig.~\ref{fig:fig2}d) and $\rm{K}$-$\rm{M}$-$\rm{K}$ (Fig.~\ref{fig:fig2}e).
Clear quasiparticle peaks are observed close to $E_{\rm{F}}$ around K point (blue arrows in Figs.~\ref{fig:fig2}d and e), and strong damping effect or the lack of the long lived electrons becomes evident as $E$ moves away from $E_{\rm{F}}$ by 50~meV and beyond.
Moving from {\rm{K} to $\rm{\Gamma}$ or $\rm{\Gamma_{2nd}}$ ($\rm{\Gamma}$ in the neighboring Brillouin zone), we find that the \rev{quasiparticle} peak and the damped intensity are dispersing downward (Figs.~\ref{fig:fig2}d and e), while they remain flat between $\rm{K}$ and $\rm{M}$ points (Fig.~\ref{fig:fig2}e).}
These photoelectron distributions as well as the one along $\rm{H}$-$\rm{K}$-$\rm{H}$ line can be explained by the theoretical band dispersions with strong band renormalization by a factor of 5 (red and yellow bold lines in Figs.~\ref{fig:fig2}c, d and e, Supplementary Note~2 and Figure~S2 \rev{and S3}).
This strong renormalization as well as the damping effects emphasize the strong electron correlation due to the localized 3$d$ orbitals of Mn atoms (Supplementary Note~3 and Figure~S\rev{4}). 

According to the theory, the Weyl points on the $\rm{K}$-$\rm{M}$-$\rm{K}$ line are located slightly above $E_{\rm{F}}$ (Fig.~\ref{fig:fig1}d).
Here we show the ARPES images observed at 60 K along several $k_{x}$ cuts (Fig.~\ref{fig:fig2}{j} inset) around $\rm{K}$ and $\rm{M}$ points in Figs.~\ref{fig:fig2}f-i obtained before (left) and after (right) dividing the intensities by the Fermi-Dirac (FD) distribution function to detect thermally populated bands above $E_{\rm{F}}$. 
In Fig.~\ref{fig:fig2}i, we find two strong intensity regions at around $\rm{M}$ point just above $E_{\rm{F}}$ and $\sim$ 40 meV below $E_{\rm{F}}$.
With increasing $k_{y}$ from $-0.80$~\AA$^{-1}$ (Fig.~\ref{fig:fig2}i), the two regions become closer in energy (Fig.~\ref{fig:fig2}h) and separated again (Figs.~\ref{fig:fig2}g \& \ref{fig:fig2}f).
They well capture what DFT calculation predicts; the flat electron (red line) and hole bands (blue line) approach each other in energy (from Fig.~\ref{fig:fig2}i to Fig.~\ref{fig:fig2}h) and finally cross to form the Weyl points (Fig.~\ref{fig:fig2}g) and separate (Fig.~\ref{fig:fig2}f) again.
In particular, this evolution of the electron band can be traced in the momentum distribution curves (MDCs) at 8~meV above $E_{\rm{F}}$ as shown in Figs.~\ref{fig:fig2}j-m.
When the $k_{x}$ cut is away from \rm{K}-\rm{M}-\rm{K} line, several anomalies are identified in the MDCs (red bars in Figs.~\ref{fig:fig2}j, l \& m). These are roughly symmetric under inversion ($k_{x} \rightarrow -k_{x}$) including the peak at $k_{x} = 0$ and two peaks/shoulders at $k_{x} \sim \pm 0.5$ ~\AA$^{-1}$. 
These should correspond to the electron bands that are weakly dispersive around K and M. Such flat parts of the bands generally cause high intensity as seen also for the flat hole band below $E_{\rm{F}}$ (blue lines in Figs.~\ref{fig:fig2}f-i).

In Fig.~\ref{fig:fig2}k, on the other hand, we observe additional anomalies arising from the crossings between the electron and hole bands along the $\rm{K}$-$\rm{M}$-$\rm{K}$ line (Fig.~\ref{fig:fig2}g).
Comparing with theory, we note that the peak (red bar) at $k_{x}$ $\sim 0.3$~\AA$^{-1}$ ($ -0.3$~\AA$^{-1}$) between $\rm{K}$ and $\rm{M}$ points most likely comes from the dispersion in the immediate vicinity of the Weyl point $W_1^{+}$ ($W_2^{-}$) (Fig. 1c), which should be located at $\sim 8$ meV above $E_{\rm F}$, given the strong band renormalization.
The peak (red bar) at  $k_{x} \sim 0.5$~\AA$^{-1}$ ($-0.5$~\AA$^{-1}$ ) between $\rm{K}$ and $\rm{\Gamma_{2nd}}$ points corresponds to a large electron band, which crosses with another band and form the Weyl point $W_1^{-} $ ($W_2^{+}$) of different chirality.
The single peak at $k_x \sim$ 0.1~\AA$^{-1}$  (blue bar in Fig.~\ref{fig:fig2}k) has been shifted from $k_{x} = 0$ by intensity gradient and would arise from the flat hole band (blue curve in Fig.~\ref{fig:fig2}g) centered at {\rm{M}} located at $\sim 14$ meV above $E_{\rm F}$.
These results appear to be consistent with the theoretically predicted Fermi surface and quasiparticle band structures.
However, strong correlation effects prohibit observation of clear spectroscopic signature of Weyl points and surface Fermi arcs (Supplementary Note 4 and Figure~S\rev{5}).
We therefore will focus on the transport properties governed by the topological nature of Weyl fermions.


Since Mn$_3$Sn is already known to exhibit a large AHE\cite{Nakatsuji15Nature} (Supplementary Note~5 and  Figure~S\rev{7}, Methods), the observation of chiral anomaly will be strong evidence for the underlying magnetic Weyl fermions. The chiral anomaly describes violation of separate number conservation laws for the left and right handed Weyl fermions in the presence of parallel electric and magnetic fields. When the resulting number imbalance between the Weyl fermions of opposite chirality is relaxed by scattering between two Weyl points, one can obtain a positive LMC\cite{Nielsen83plb,Son2013,Goswami2015}, while the TMC remains negative. Very recently, such anisotropic magnetoconductance due to chiral anomaly has been experimentally confirmed in the weakly correlated materials such as the Dirac semimetal Na$_3$Bi, the Weyl semimetal TaAs, and the quadratic band touching system GdPtBi\cite{Xiong2015,Huang2015,Zhang2016,Hirschberger16naturemat}. 

To identify the effects of chiral anomaly, we prepared single crystals with more Mn (Mn$_{3.06}$Sn$_{0.94}$) than crystals used for ARPES so that $E_{\rm F}$ becomes closer to the Weyl points (Figs.~\ref{fig:fig1}b \& d). 
Taking account of the strong band renormalization, we can estimate the energy separation between the Weyl points and $E_{\rm F}$ to be $\sim 5$ meV, which is close to the TaAs case\cite{Zhang2016}.
The magnetoconductance was measured in a schematic configuration shown in the insets of Fig.~\ref{fig:fig3}a (Fig.~\ref{fig:fig3}b) where the current $I$ is applied along the [01$\bar{1}$0] ([2$\bar{1} \bar{1}$0]) axis, and $B$ is either parallel (red) or perpendicular (blue) to $I$. Strongly anisotropic magnetoconductance was observed at all $T$s measured and typical results at 60 K are shown in Figs.~\ref{fig:fig3}a \& b. When $B$ is applied parallel (perpendicular) to $I$, the longitudinal conductance increases (decreases) with $B$. Moreover, the size of the positive magnetoconductivity, $\Delta \sigma (B)$, becomes significantly larger at lower $T$s (Figs.~\ref{fig:fig3}c \& d). When $B \parallel I$, the magnetoconductivity does not saturate and keeps increasing with $B$. 

To clarify the anisotropic character, we scanned the angle $\theta$ between $B$ and $I$ directions for the magnetoconductance measurements under a field of 9 T, which is much higher than the coercivity ($\sim$200 G) in the magnetization curve, and thus the measurements were all performed in a single domain state. Figures 3e and 3f show the angle dependence of the relative magnetoconductances $\Delta \sigma({\rm 9~T})=\sigma({\rm 9~T})-\sigma({\rm 0~T})$ at selected temperatures. Here, the positive and negative $\Delta \sigma(\theta)$ indicate the positive and the negative magnetoconductance, respectively. 

A very sharp angle dependence is found; the positive magnetoconductance becomes maximized when $B$ is set exactly parallel to $I$ ($\theta = 0 ^{\circ}$). A small change in the angle $\theta$ from $\theta = 0 ^{\circ}$ leads to a very sharp but symmetric decrease to a negative value. This type of a sharp angle dependence has never been seen for the magnetoconductance in magnetically ordered states. For example, in ferromagnets, the anisotropic magnetoconductance rather gradually depends on $\theta$ \cite{McGuire}. 
Actually, the positive magnetoconductance with such a very narrow window near $ B \parallel I$ is very similar to the previous experimental observations of chiral anomaly \cite{Xiong2015,Huang2015,Zhang2016,Hirschberger16naturemat}. In addition, the positive magnetoconductance associated with the chiral anomaly is expected to be more significant at low temperatures when the inelastic scattering effects decrease. Indeed, the significantly anisotropic positive magnetoconductance becomes more enhanced as we cool down the system. 
Moreover, when the Fermi level is tuned to be further away from the Weyl points, as in the crystals used for ARPES (Mn$_{3.03}$Sn$_{0.97}$), we have observed that both AHE and LMC are reduced while TMC remained almost unaffected (Supplementary Figure S\rev{7}).
Such properties of anisotropic magnetoconductance for samples with comparable strength of magnetic fluctuations cannot be explained in terms of the field-induced suppression of magnetic fluctuations and weak localization (Supplementary Notes~6\&7 and  Figure~S\rev{7}).
On the other hand, it is known that the positive LMC can also arise from the inhomogeneous current distribution (current jetting) in the sample \cite{Pippard}, which we rule out by carefully studying the contact and sample dependence of our data (Supplementary Note~7 and  Figure~S\rev{8}). 
Finally, recent semiclassical calculations indicate that the type-II Weyl fermions show a $B$-linear positive magnetoconductance, in comparison with the quadratic dependence expected for type-I Weyl semimetals\cite{Zyuzin2016}. 
Thus, the observed $B$-linear increase of the positive magnetoconductance at low field regime (Figs.~3a-3d) is consistent with chiral anomaly of the type-II Weyl fermion state predicted for Mn$_3$Sn\cite{Yang2017njp}.

Magnetic Weyl fermions in a correlated metal may provide many scientific and technological possibilities. Their availability at room temperature allows further in-depth study of emergent phenomena involving large Berry curvatures. Since the magnetic structure and the domain wall textures for such a ``Weyl magnet" can be easily manipulated
	with a magnetic field, one can also control the location of Weyl nodes and the Berry vector potential in a space and time dependent manner, which can lead to emergent electromagnetism and spin/electric current generation\cite{Landsteiner2014}. Our experimental observations thus initiate the basic research on correlated magnetic Weyl fermions, which may well lead to novel electronic and spintronic technology for future applications.

\bibliography{Nature_rev0603}

\subsection{Methods\\}
{\bf{DFT calculation.}}
Electronic structure calculations were performed for the antiferromagnetic state of Mn$_3$Sn with the QUANTUM ESPRESSO package\cite{Giannozzi2009}, using a relativistic version of ultrasoft pseudo potentials.
We use the lattice constants obtained by the powder X-ray measurements at 60 K and the Wyckoff position of the Mn
6$h$ atomic sites $x=$0.8388 from the literature~\cite{Tomiyoshi1982,Brown90jpcm}.
All of the calculated results were obtained for the magnetic texture shown in Supplementary Figure~S1a without a perturbation of the weak ferromagnetic moment.
\\
{\bf{Mn$_3$Sn single crystal.}}
Single crystals used in this work were grown by Czochralski method using a commercial tetra-arc furnace (TAC-5100, GES) or a Bridgman method using a homemade furnace.
Our single crystal and powder X-ray measurements find the single phase of hexagonal Mn$_3$Sn with room temperature lattice constants of $a = 5.687(5.662)$ \AA~and $c = 4.548 (4.529)$ \AA~for crystals used for ARPES (transport) measurements.
Analysis using inductively coupled plasma (ICP) spectroscopy found that the composition of the single crystals used for ARPES and transport measurements are slightly off-stoichiometric, Mn$_{3.03}$Sn$_{0.97}$, and Mn$_{3.06}$Sn$_{0.94}$, respectively.

The extra Mn induces the electron doping.
In our first-principles calculation for Mn$_{3}$Sn, the number of the Mn-$s$, Mn-$d$, Sn-$s$, Sn-$p$ electrons are estimated to be 0.84, 5.85, 1.67, and 3.26, respectively.
For Mn$_{3.06}$Sn$_{0.94}$, it is reasonable to assume that the electron occupancy of the Sn-$s$, Sn-$p$ and Mn-$s$ orbital are the same as those of Mn$_{3}$Sn, since the low-energy states around the
Fermi level are mainly formed by the Mn-$d$ orbital (Supplementary Note~3 and Figure~S\rev{4}).
Then the number of electrons doped into the Mn-$d$ orbital is estimated to be 0.024, which shifts the Fermi energy ($E_{\rm{F}}$) by 0.04~eV for Mn$_{3.06}$Sn$_{0.94}$ in comparison with the stoichiometric Mn$_3$Sn case, as shown in Figs.~1b \& 1d.
With a similar calculation, the shift of $E_{\rm{F}}$ is estimated to be 0.019 eV for Mn$_{3.03}$Sn$_{0.97}$.
This provides better consistency with our ARPES data, and we use the shifted $E_{\rm{F}}$ to compare our ARPES results with theory.
In addition, our ARPES measurements confirm the electron doping effect of the extra Mn in off-stoichiometric samples, which shows quantitative agreement with the theoretical estimate (Supplementary Note \rev{8} and Figure~S\rev{9}).
Unless it is necessary to specify the exact composition of the crystals used for measurements, we use ``Mn$_3$Sn'' to refer to our crystals for clarity throughout the paper.
\\
{\bf{ARPES set-up.}} 
ARPES experiment has been performed at BL28 of Photon factory (PF) and at 1$^{2}$ beamline of BESSY-II.
Various photon energy ($h\nu$) were used ranging from 50 to 150~eV from synchrotron radiation. 
The photoelectrons were acquired by hemispherical analyzers, ScientaOmicron SES-2002 (PF) and R8000 (BESSY-II). 
The overall energy resolution was set to about 20 meV at 103~eV, and thus the total Fermi-edge broadening was estimated to be $\sim 30$~meV at the measurement temperature.
We also performed soft X-ray ARPES (SX-ARPES) measurement at BL25SU at SPring-8.
The total experimental energy resolution of the SX-ARPES was set to about 45~meV at $h\nu$ of 330~eV and the photoelectrons were acquired by ScientaOmicron DA30. 
In all ARPES experiments, angular resolutions are 0.3~degrees.
Samples with the composition Mn$_{3.03}$Sn$_{0.97}$ were cleaved at approximately 60~K, exposing flat surfaces corresponding to the (0001) cleavage plane.
The sample temperature was kept at 60~K during the measurement to avoid a cluster glass phase with ferromagnetic moment which appears below 50~K \cite{Nakatsuji15Nature}.
To prepare a single magnetic domain, we applied a field of 2000 Oe along the $x$-axis at room temperature before cleaving the crystals.
\\
{\bf{Transport and magnetization measurements.}}
For the transport measurements, single crystals with the composition Mn$_{3.06}$Sn$_{0.94}$ were shaped into a rectangular sample (typical size: $\sim 0.1 \times 1.5 \times 1.5$ mm$^3$). The longitudinal and Hall resistivity measurements were made using the four-probe method. Both current and voltage terminals were spot-welded on the polished sample surface using the Au-wires. For current terminals, all the area of the sides was painted using Ag paste to avoid inhomogeneous distribution of the electric field $E$. 
To rotate a sample with respect to the magnetic field, the Quantum Design Horizontal Rotator option for the Physical Property Measurement System (PPMS) was used. The rotation angle for the uniaxial rotator can continuously move with angular step size of 1 degree.
The temperature dependence of the magnetization was measured using a commercial magnetometer (MPMS, Quantum Design). The measurement above 350 K was made using the oven option.

%

\begin{addendum}
	\item
	This work was supported by CREST, Japan Science and Technology Agency, Grants-in-Aid for Scientific Research (Grant Nos. 16H02209, 25707030), by Grants-in-Aid for Scientific Research on Innovative Areas ``J-Physics" (Grant Nos. 15H05882 and 15H05883), and ''Topological Materials Science" (Grant No.16H00979) and Program for Advancing Strategic International Networks to Accelerate the Circulation of Talented Researchers (Grant No. R2604) from the Japanese Society for the Promotion of Science. P. G. was supported by JQI-NSF-PFC and LPS-MPO-CMTC. The use of the facilities
	of the Materials Design and Characterization Laboratory at the Institute for Solid State Physics, The University of Tokyo, is gratefully acknowledged.
	
	\item[Author Contributions] 
	S.N. planned the experimental project.
	K.K., C.B., conducted ARPES experiment and analyzed the data.
	M.N., S.A., R.N., S.S., T.Kon., N.I., K.O., H.K., T.M., A.V. supported ARPES experiment.
	A.N., I.M., S.N. made Mn$_3$Sn single crystals. R.I. performed chemical analyses.
	T.T., I.M., S.N. performed transport experiments and collected data.
	P.G. provided theoretical insight.
	M.S., T.Kor., M.O., R.A. calculated the band structure.
	K.K., T.T., P.G., R.A., S.N. wrote the paper; 
	All authors discussed the results and commented on the manuscript. 
	
	\item[Competing Interests] The authors declare that they have no competing financial interests.
	
	\item[Correspondence] Correspondence and requests for materials
	should be addressed to S.N.~(email: satoru@issp.u-tokyo.ac.jp).
\end{addendum}
\begin{figure}

	\caption{
			{\bf Magnetic structure, and three-dimensional bulk band dispersion in Mn$_{3}$Sn.}
			{\bf a}, Magnetic texture in the kagome lattice. Here, we take the $x$, $y$, and $z$ coordinates along $[2\bar{1}\bar{1}0]$, $[01\bar{1}0]$, and $[0001]$.
			Mn moments ($\sim$~3~$\mu_{\rm{B}}$, $\mu_{\rm{B}}$ is the Bohr magneton) lying in each $x$-$y$ kagome plane form a 120$^{\circ}$ degree structure. This magnetic structure allows spin canting, leading to a very small net in-plane magnetization of 0.002 $\mu_{\rm{B}}$ per Mn. Arrows indicate Mn moments which have the local easy-axis parallel to the in-plane direction along the $x$ axis. 
			{\bf b}, Overview of the bulk band structure calculated by DFT for Mn$_{3}$Sn along high symmetry lines. In comparison with the stoichiometric Mn$_3$Sn, crystals with the composition Mn$_{3.03}$Sn$_{0.97}$ (Mn$_{3.06}$Sn$_{0.94}$) used for ARPES (transport) measurements are doped with more conduction electrons due to extra Mn, and have a higher Fermi energy $E_{\rm{F}}$, as indicated by the red (blue) dotted line (Methods). 
			{\bf c}, Distribution of the Weyl points in the bands on $k_{x}$-$k_{y}$ plane at $k_{z}$=0 near $E_{\rm{F}}$ for the magnetic texture shown in {\bf a}.
			Two pairs of the Weyl nodes with different chirality (W$^{+}$, W$^{-}$) are shown by open and solid circles. The dotted circles schematically show the hypothetical nodal rings that appear when SOC is turned off.
			{\bf d}, Enlarged DFT band structure around $\rm{M}$ and $\rm{M}'$ points cut along distinct high symmetry lines (red and blue arrows in {\bf c}, respectively).
			The Weyl (band-crossing) points with opposite chirality are denoted by open and solid arrows, respectively.			
			\label{fig:fig1}  
		}

\end{figure}
\begin{figure}

		\caption{
			{\bf ARPES band mapping near the Fermi level compared to DFT band calculation for Mn$_3$Sn.}
			{\bf a}, Brillouin zone, showing the momentum sheet at $k_{z}=0$ and different high symmetry lines, for which ARPES data were taken.
			{\bf b}, ARPES intensity at $E_{\rm{F}}$ in the $k_x$-$k_y$ plane crossing $\Gamma$ by using $h\nu = 103$~eV and the calculated Fermi surface (purple curves).
			The dashed line indicates the hexagonal Brillouin zone.
			Taking account of the electron doping in the ARPES sample Mn$_{3.03}$Sn$_{0.97}$, here we use the Fermi energy indicated by the red line in Fig.~\ref{fig:fig1}b.
			{\bf c}, Left, $k_z$ dispersion along $\rm{H}$-$\rm{K}$-$\rm{H}$ high symmetry line (black arrows in {\bf a}).
	Right, the corresponding energy distribution curves (EDCs).
			The bold lines indicate EDC cuts at high symmetry {\bf{k}} points.
			The ARPES maps are compared to the band calculations with strong band renormalization by a factor of 5 (white lines).
	The red bold lines indicate the specific theoretical bands which likely dominate spectral intensities (Supplementary Note~2 and Figure~S2 \rev{and S3}).
{\bf d-e}, ARPES band mapping along the high symmetry lines, {\bf d}, $\rm{K}$-$\rm{\Gamma}$-$\rm{K}$ (red arrows in {\bf a}) and {\bf e}, $\rm{K}$-$\rm{M}$-$\rm{K}$ (blue arrows in {\bf a}), in the $k_{x}$-$k_{y}$ plane at $k_{z}=0$ (left), and their EDCs (right).
			The yellow bold lines indicate the specific theoretical bands which likely dominate spectral intensities (Supplementary Note~2 and Figure~S2).
		{\bf f-i}, ARPES $E$-$k_{x}$ cuts and the theoretical band structures at different $k_{y}$ as shown in the inset of {\bf j}.
{\bf j-m}, Corresponding momentum distribution curves (MDCs) at $E$-$E_{\rm{F}} \sim$ 8 meV.
			In the ARPES intensity maps in {\bf f-i}, we compare (left) the original ARPES intensity and (right) the intensities divided by the energy-resolution convoluted Fermi-Dirac (FD) function at the measured temperature of 60~K to remove the cut-off effect near $E_{\rm{F}}$.
			The electron and hole bands responsible for forming the Weyl nodes are highlighted with red and blue lines, respectively.
			The yellow bold lines correspond to the band dispersions highlighted in {\bf d, e}. 
			The anomalies in their MDCs are marked by colored bars.
Notably, these anomalies are intrinsic as the FD-division process does not affect the MDC shape.
			\label{fig:fig2}
		}

\end{figure}
\begin{figure}

		\caption{
		{\bf Strongly anisotropic magnetoconductance in Mn$_3$Sn.}
		{\bf a-b,} Magnetic field dependences of the magnetoconductivity $\Delta \sigma(B)= (\sigma(B)-\sigma(0))$ at 60 K for $I \parallel B$ (red) and $I \perp B$ (blue). The dotted lines indicate linear fits.
		Inset, configurations used for the magnetoconductance measurements. 
		The green and red arrows respectively indicate the magnetic field $B$ and electric current $I\parallel E$. The red planes indicate the $x$-$y$ (kagome) plane in Mn$_3$Sn.
		{\bf c-d,} Magnetic field dependence of the magnetoconductivity $\Delta \sigma(B)= (\sigma(B)-\sigma(0))$ at various temperatures for {\bf c}, $I \parallel B\parallel$ [01$\bar{1}$0], {\bf d}, $I \parallel B\parallel$ [2$\bar{1}$$\bar{1}$0].  The temperature dependence of $\sigma(0)$ is given in Supplementary Figure~S\rev{6}.
Inset, results at 300 K. {\bf e-f}, Angle dependence of the magnetoconductivity $\Delta \sigma(\theta)$ at 9 T measured at various temperatures with  {\bf e}, $I \parallel$ [01$\bar{1}$0], {\bf f}, $I \parallel$ [2$\bar{1}$$\bar{1}$0]. Inset, configurations with the rotation axis along {\bf e}, [0001] and {\bf f}, [01$\bar{1}$0]. $\theta$ is defined by the angle between $B$ and $I$ directions. 
					\label{fig:fig3}
		}

\end{figure}

	\pagebreak
\begin{figure}
	\begin{center}
		\hspace{-1.4cm}
		\includegraphics[width=7in]{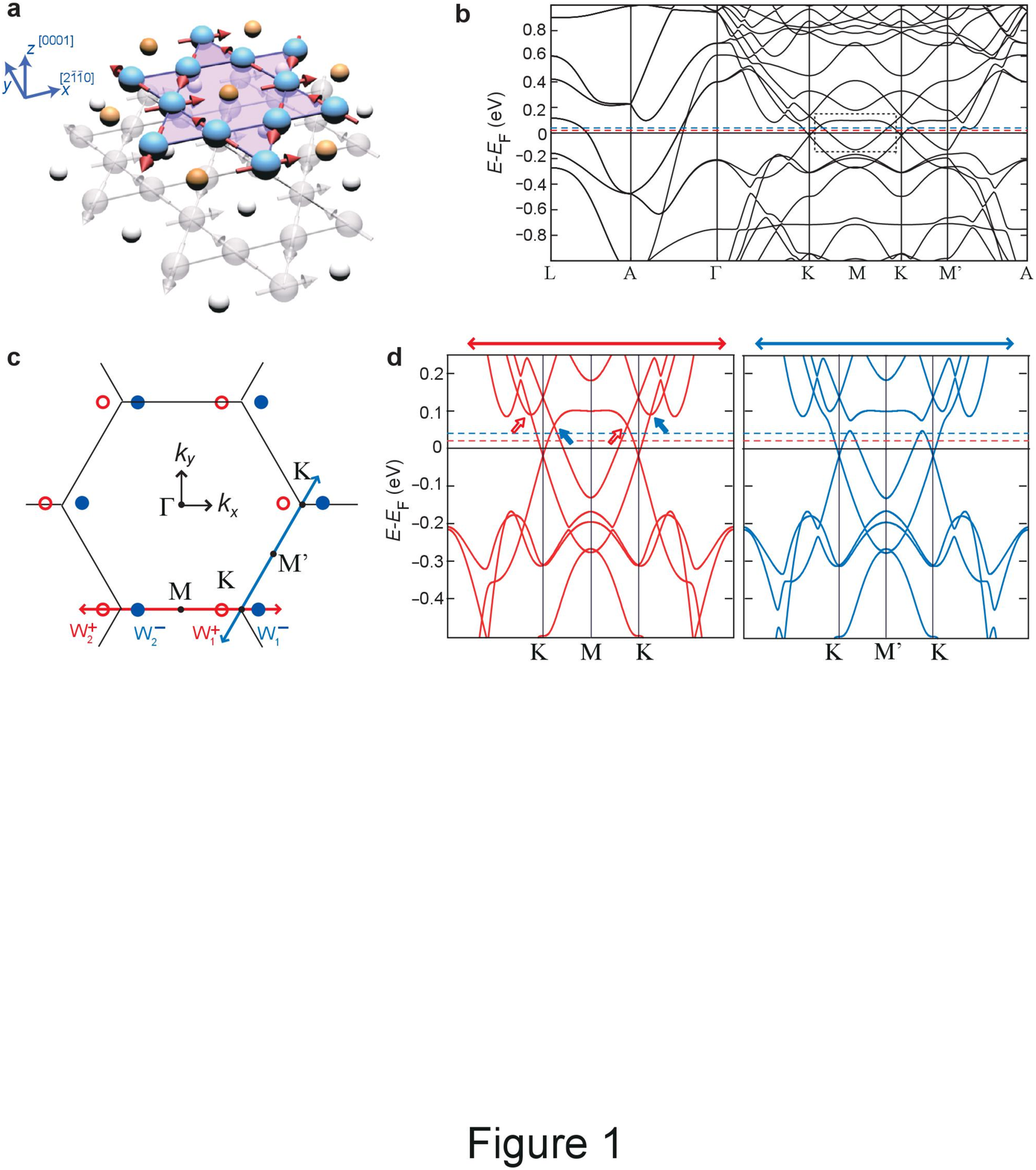}
	\end{center}
\end{figure}

\begin{figure}
	\begin{center}
		\hspace{-1.4cm}
		\includegraphics[width=7in]{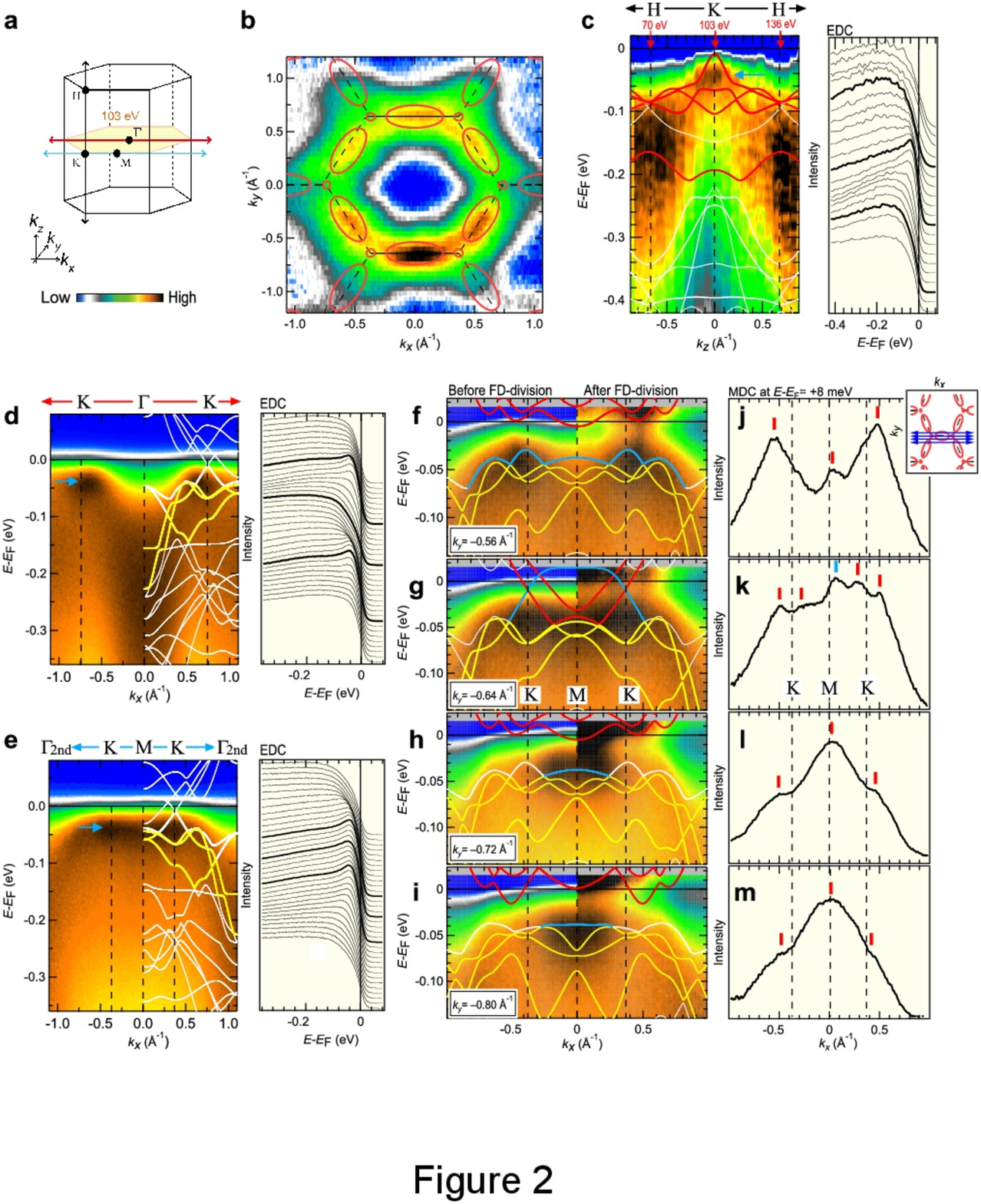}
	\end{center}
\end{figure}

\begin{figure}
	\begin{center}
		\hspace{-1.4cm}
		\includegraphics[width=7in]{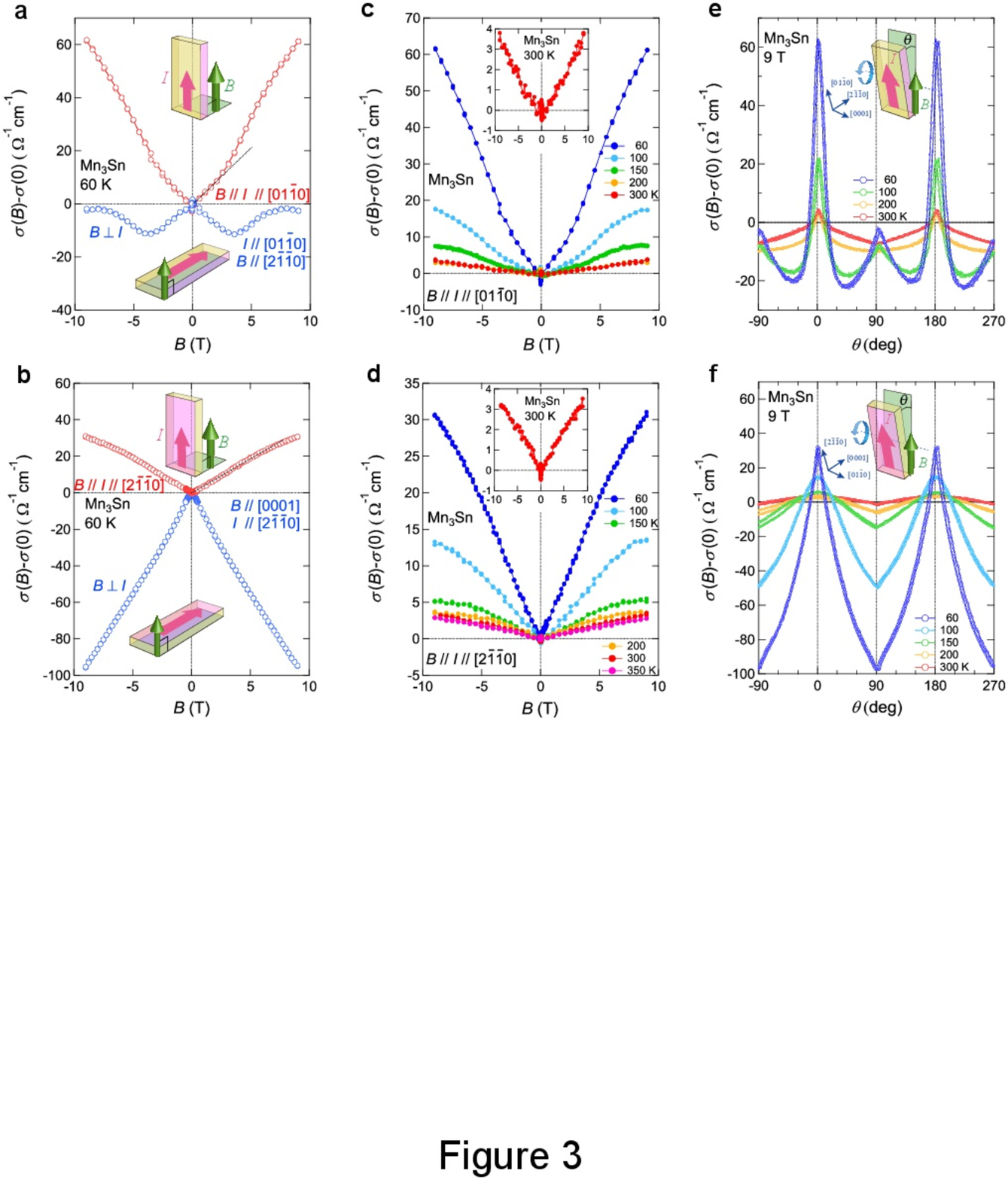}
	\end{center}
\end{figure}

\newpage
\begin{center}
\title{\bf \Large{Supplementary Information: \\Evidence for Magnetic Weyl Fermions in a Correlated}
	Metal}
\end{center}
\noindent
\author{K.~Kuroda$^{1}$\footnotemark[2]}
\author{T.~Tomita$^{1,3}$\footnotemark[2]}
\author{M.-T.~Suzuki$^{2,3}$}
\author{C.~Bareille$^1$}
\author{A.~A.~Nugroho$^{1, 4}$}
\author{P.~Goswami$^5$}
\author{M.~Ochi$^6$}
\author{M.~Ikhlas$^{1,3}$}
\author{M.~Nakayama$^1$}
\author{S.~Akebi$^1$}
\author{R.~Noguchi$^1$} 
\author{R.~Ishii$^1$}
\author{N.~Inami$^7$}
\author{K.~Ono$^7$}
\author{H.~Kumigashira$^7$}
\author{A. Varykhalov$^8$}
\author{T.~Muro$^9$}
\author{T.~Koretsune$^{2,3}$}
\author{R.~Arita$^{2,3}$}
\author{S.~Shin$^1$}
\author{Takeshi~Kondo$^1$}
\author{and S.~Nakatsuji$^{1,3*}$}

\maketitle
\small{
\begin{affiliations}
    	\item \footnotesize{Institute for Solid State Physics, University of Tokyo, Kashiwa 277-8581, Japan}
		\item \footnotesize{RIKEN Center for Emergent Matter Science (CEMS), 2-1 Hirosawa, Wako, Saitama 351-0198, Japan}
		\item \footnotesize{CREST, Japan Science and Technology Agency (JST), 4-1-8 Honcho Kawaguchi, Saitama 332-0012, Japan}
		\item \footnotesize{Faculty of Mathematics and Natural Sciences, Institut Teknologi Bandung, Jl. Ganesha 10, 40132 Bandung, Indonesia}
		\item \footnotesize{Condensed Matter Theory Center and Joint Quantum Institute, Department of Physics, University of Maryland, College Park, Maryland 20742- 4111 USA}
	\item \footnotesize{Department of Physics, Osaka University, Machikaneyama-cho, Toyonaka, Osaka 560-0043, Japan}
	\item \footnotesize{Institute of Materials Structure Science, High Energy Accelerator Research Organization (KEK), Tsukuba, Ibaraki 305-0801, Japan}	
	\item \footnotesize{Helmholtz-Zentrum Berlin f{\"u}r Materialien und Energie, Elektronenspeicherring BESSY II, Albert-Einstein-Strasse 15, 12489 Berlin, Germany	}
	\item \footnotesize{Japan Synchrotron Radiation Research Institute (JASRI), 1-1-1 Kouto, Sayo, Hyogo 679-5198, Japan}
	
\end{affiliations}
}
%
\subsection{Supplementary Note 1 : Magnetic structures in Mn$_3$Sn.}
In Mn$_3$Sn, the Mn moments of approximately 3$\mu_{\rm{B}}$, lying in the $x$-$y$ kagome plane, form a non-collinear 120$^{\circ}$ spin ordering below the N{\'{e}}el temperature of 430 K~\cite{Nakatsuji15Nature}. Unlike the normal 120$^{\circ}$ order seen in geometrically frustrated magnets, it has an opposite vector chirality and thus is called inverse triangular spin configuration (Supplementary Fig.~S\ref{texture}).
Interestingly, neutron diffraction measurements and theoretical analyses have shown that the inverse triangular spin structure should have no in-plane anisotropy energy up to the fourth-order term~\cite{Nagamiya1982,Tomiyoshi1982}.
On the other hand, this spin structure has orthorhombic symmetry, which allows the spin canting in the $x$-$y$ kagome plane. 
For example, in the configuration shown in Supplementary Fig.~S\ref{texture}a, only one of the three moments is parallel to the local easy-axis along the $x$ direction~\cite{Tomiyoshi1982,Brown90jpcm}, and the canting of the other two spins towards the local easy-axis may lead to the weak ferromagnetic moment of the order of approximately 0.002 $\mu_{\rm{B}}$ per Mn atom, as observed in experiment~\cite{Nakatsuji15Nature}. 
With such a vanishingly small anisotropy energy, the coupling of the weak magnetization to the magnetic field plays an essential role for the magnetic field control of the sublattice moments.
Namely, a small coercive field $\sim 200$~Oe is enough to rotate the sublattice moments in the non-collinear antiferromagnetic structure, as observed in experiment \cite{Nakatsuji15Nature}.
Supplementary Figures~S\ref{texture}a and S\ref{texture}b show two representative magnetic configurations that are available when the field is applied along the $x$ and $y$ directions, respectively.
For both configurations, Mn$_3$Sn exhibits a large anomalous Hall effect~\cite{Nakatsuji15Nature}.

\subsection{Supplementary Note 2 : Strong correlation effect in Mn 3$d$ bands.}
In Supplementary Fig.~S\ref{Ekmap_calc}, we show the band maps obtained by using angle-resolved photoesmission spectroscopy (ARPES) with $p$-polarized light along high symmetry momentum cuts and their comparison with theoretical band dispersion renormalized by various band renormalization factors ($m^{\rm{ARPES}}/m^{\rm{DFT}}$).
The ARPES maps cut on the $k_x$-$k_y$ plane are obtained after dividing the intensities by the Fermi-Dirac distribution function (Supplementary Figs.~S\ref{Ekmap_calc}a-j).
In all momentum cuts, we observe clear quasiparticle peaks near $E_{\rm{F}}$, whose energy position is indicated by blue arrow in Supplementary Fig.~S\ref{Ekmap_calc}.
Strong damping effect or the lack of the long lived electrons becomes clear as $E$ moves away from $E_{\rm{F}}$ by 50~meV and beyond.
Moving from ${\rm{K}}$ to ${\rm{M}}$ (Supplementary Fig.~S\ref{Ekmap_calc}a), we find the flat distribution of the intensities while the peak and the damped intensities disperse downward around ${\rm{\Gamma}}$.
Without considering the band renormalization ($m^{\rm ARPES}/m^{\rm DFT}$=1), the band calculations do not match  the ARPES results (Supplementary Figs.~S\ref{Ekmap_calc}a and S\ref{Ekmap_calc}f). 
Significantly, upon increasing $m^{\rm ARPES}/m^{\rm DFT}$, several theoretical bands (yellow bold lines) become more consistent with the ARPES intensity maps.
When $m^{\rm ARPES}/m^{\rm DFT}$=5 is used, the energy position of these bands (yellow arrows) agrees well with the peak (blue arrow, Supplementary Figs.~S\ref{Ekmap_calc}e and S\ref{Ekmap_calc}j).
This further explains the downward dispersion observed around ${\rm{\Gamma}}$.
We therefore use $m^{\rm ARPES}/m^{\rm DFT}$=5 for our discussion in the main text.
Moreover, the observed $k_z$ dispersion (Supplementary Figs.~S\ref{Ekmap_calc}k-o) and the fine structures seen in the momentum distribution curves around $\rm{M}$ point (Figs.~2f-i in the main text) can be understood in accordance with the renormalized theoretical bands.
The localized 3$d$ orbitals of Mn atoms in the kagome lattice are responsible for the strong correlation effects in Mn$_3$Sn (Supplementary Note~3 and Fig.~S\ref{resonance}).
The strong renormalization by a factor of $\sim5$ is more pronounced than in other 3$d$ electron systems~\cite{Yoshida05prl,Borisenko10prl,Lev15prl}.

In Supplementary Fig.~S\ref{polarization}, we further show ARPES band maps obtained by using different light polarizations.
These APRES intensity maps are compared with calculated bands renormalized by the factor of $m^{\rm ARPES}/m^{\rm DFT}$=5.
Among the theoretical band dispersions, those showing good agreement with the ARPES data with $p$-polarized light (Supplementary Figs.~S\ref{polarization}c and S\ref{polarization}e), are highlighted by yellow bold lines.
By switching the polarization of the incident light to $s$-polarization, we find more spectral weights at high binding energy around $E$-$E_{\rm{F}}$=$-$0.2~eV, particularly for the case of Supplementary Fig.~S\ref{polarization}f compared to Supplementary Fig.~S\ref{polarization}e.
The spectral weights sensitive to $s$-polarization likely explain other theoretical bands highlighted by the sky-blue bold lines as shown in Supplementary Figs.~S\ref{polarization}d and S\ref{polarization}f.

\subsection{Supplementary Note 3 : Results of soft X-ray angle-resolved photoemission spectroscopy.}
Since the spectral weight in ARPES with vacuum ultraviolet is generally governed by surface signal, we performed more bulk-sensitive soft X-ray ARPES (SX-ARPES).
In the photoelectron distribution at $E_{\rm{F}}$ on the wide $k_x$-$k_y$ sheet at $k_z$=0 in the 7th Brillouin zone ($h\nu$=330~eV; Supplementary Fig.~S\ref{resonance}a), we see the strong photoelectron intensity only around zone boundary as observed by using $h\nu$ of 103~eV (Fig.~2b in the main text).
This result strongly supports our conclusion that ARPES with $h\nu$ of 103~eV detects the bulk electronic structures. 

In Supplementary Fig.~S\ref{resonance}b, we summarize the photoelectron intensity near $E_{\rm{F}}$ as a function of $h\nu$. 
The intensity jumps at $h\nu\sim$640 eV and 652 eV, corresponding to the energies associated with Mn 2$p$-Mn 3$d$ excitations.
The variation of the peak intensity with $h\nu$ can be described by a Fano line shape as expected for the resonance behavior, which is a consequence of a localized character of the Mn 3$d$ states.
From this resonant photoemission, we unambiguously show that the electronic structure near $E_{\rm{F}}$ is predominantly formed by Mn 3$d$ orbitals, which is consistent with theory as shown in Supplementary Fig.~S\ref{resonance}c.
Therefore, the Weyl fermion states in Mn$_3$Sn should be predominantly formed by the Mn 3$d$ orbitals.

\subsection{Supplementary Note 4 : Fermi-arc surface states in (0001) surface of Mn$_3$Sn with DFT.}
In our DFT calculation, the Fermi-arc surface-states connecting the pair of the Weyl nodes are observed around $\rm{K}$ as shown in Supplementary Fig.~S\ref{arc}a, where we plot the surface weight difference of wave functions among top and bottom surfaces at 59~meV above the theoretical 
$E_{\rm{F}}$ for the 50-layer (0001) slab model constructed using the bulk transfer integrals.
Red and blue colours indicate the Fermi arcs on the top and bottom surfaces, respectively.
Since the orthorhombic symmetry of the magnetic spin structure determines the momentum-location of the magnetic Weyl nodes, the Fermi arcs are very different from those reported for the different magnetic texture in the literature~\cite{Yang2017njp}.
Note that residual bulk states have a large projection except for a small $k$-region around $\rm{K}$ points where the Fermi-arc surface states are located.
At a slightly lower energy at 19~meV above the theoretical $E_{\rm{F}}$, the projection gap becomes even smaller and the surface states form a resonance with the bulk states (Supplementary Fig.~S\ref{arc}b).
The surface Fermi arcs are thus confined only in a small $E$-$k$ region in theory, which is moreover renormalized by the strong correlation effect of Mn 3$d$ electrons as we demonstrated by ARPES.
These situations should make it very difficult to identify the Fermi-arc state under strong correlation effect of Mn$_3$Sn by using ARPES.
In the main text, we therefore presented the transport properties governed by the topological nature as evidence for Weyl fermions in the strong correlated magnet.

\subsection{Supplementary Note 5 : Anomalous Hall Effect.}
As reported previously\cite{Nakatsuji15Nature}, Mn$_3$Sn exhibits a large anomalous Hall effect (AHE). We confirmed that the crystals used in this study show a large AHE.
Supplementary Figures~S\ref{Hall}d and S\ref{Hall}e show the field dependence of the Hall conductivity $\sigma_{\rm H} = \sigma_{zy}\approx -\rho_{zy}/(\rho_{zz}\rho_{yy})$ for Mn$_{3.03}$Sn$_{0.97}$ and Mn$_{3.06}$Sn$_{0.94}$. Here, $\rho_{zy}$ is the Hall resistivity measured under magnetic field $B$ along  [2$\bar{1}\bar{1}$0] with electrical current $I$ along [01$\bar{1}$0]. $\rho_{zz}$ and $\rho_{yy}$ are the longitudinal resistivity measured with electrical current $I$ along [0001] and [01$\bar{1}$0], respectively. $\sigma_{\rm H}$ of Mn$_{3.03}$Sn$_{0.97}$ (Mn$_{3.06}$Sn$_{0.94}$) shows a hysteresis with a coercivity of $\sim 200$~Oe with a spontaneous zero-field value of $\sim40 (42)~\Omega^{-1}$cm$^{-1}$ at 300 K and $\sim110 (130)~ \Omega^{-1}$cm$^{-1}$ at 100 K.
In large field $B$, the size decreases with a negative slope as a function of $B$.

\subsection{Supplementary Note 6 : Sample dependence of the Negative Longitudinal Magnetoresistance}
Here, we compare the magnetic and transport properties for two different crystals used mainly for ARPES and magnetoresistance measurements, respectively.
We note that Mn concentration is different only by 1\% between the two crystals, namely, Mn$_{3.03}$Sn$_{0.97}$ (ARPES) and Mn$_{3.06}$Sn$_{0.94}$ (Magnetoresistance). 
As expected, both have nearly the same temperature dependence of the magnetization and N{\'{e}}el temperatures (Supplementary Fig.~S\ref{Hall}a), indicating the comparable strength of magnetic fluctuations in both samples. However, the two samples have different distance $\Delta = E_{\rm{F}} -W$ between the Fermi energy $E_F$ and the reference energy $W$ for the Weyl point. As a result, the longitudinal magnetoconductance is significantly reduced by $\sim60$\% as the Fermi energy moves away from the Weyl point from $\Delta \approx 5$ meV (Mn$_{3.06}$Sn$_{0.94}$) to 8 meV (Mn$_{3.03}$Sn$_{0.97}$), while the transverse magnetoconductance remains almost unaffected (Supplementary Figs.~S\ref{Hall}b and S\ref{Hall}c). In the weak magnetic field regime, the semi-classical calculations suggest that the chiral anomaly-induced positive longitudinal magnetoconductance is inversely proportional to $\Delta^2$~\cite{Son2013}.  Our observation of $\sim60$\% reduction is consistent with the theoretical expectation. In addition, the AHE becomes smaller with increasing $\Delta$ (Supplementary Figs.~S\ref{Hall}d and S\ref{Hall}e, Supplementary Note~5). The strong anisotropic reduction of the magnetoconductance cannot be attributed to the field-induced suppression of spin fluctuations. Instead, the reduction in both longitudinal magnetoconductance and AHE is consistent with Berry curvature effects of the Weyl points\cite{Son2013,Goswami2013}.

\subsection{Supplementary Note 7 : Origins of the Negative Longitudinal Magnetoresistance (LMR).}
A solid state system can possess negative LMR due to several reasons such as (i) field induced suppression of scattering from magnetic impurities, (ii) inhomogeneous current distribution or current jetting effect, (iii) weak localization effects and (iv) chiral anomaly of Weyl fermions. Below we provide arguments to rule out the first three conventional mechanisms as the possible source of concomitantly observed negative LMR and positive transverse magnetoresistance (TMR) in Mn$_3$Sn. This helps us to establish chiral anomaly of Weyl fermions as the intrinsic mechanism behind the observed magnetoresistance properties.

When electronic scattering rate from local moments and magnetic impurities is suppressed by an applied external magnetic field, the transport lifetime is enhanced, leading to negative magnetoresistance along all directions (i.e, both LMR and TMR become negative). This is routinely observed in heavy fermion compounds above the Kondo temperature scale (a regime where $f$-electrons are not a part of the coherent excitations defined around a large Fermi surface and conduction electrons are thus scattered by local moments). Concomitantly, negative LMR and TMR are also observed in ferromagnetic compounds~\cite{Pippard,Coleman}, when an applied magnetic field suppresses fluctuations of the ferromagnetic order parameter (thus decreasing the inelastic scattering rate for conduction electrons). Therefore, observed negative LMR and positive TMR in Mn$_3$Sn cannot be explained on the basis of field induced suppression of magnetic scattering. 

In materials with high carrier mobility and strong resistance anisotropy (quantified by the ratio $A$=TMR/LMR), an inhomogeneous distribution of the current flowing inside the sample can give rise to negative LMR. The negative LMR arising due to current-jetting effect shows strong dependence on sample geometry/size \cite{Pippard, Hu2005, Reis2016}. We first note that Mn$_3$Sn does not possess large carrier mobility like weakly correlated semimetals and zero-gap semiconductors. In addition, the ratio $A$ remains of the order of unity for the entire range of applied magnetic field strengths. Thus, we do not expect any significant current-jetting effect. Nevertheless, we have carried out explicit measurements to rule out the current-jetting effect. Supplementary Figure S\ref{contact}a shows our experimental set up for the sample.
The sample is prepared with six spot-welded voltage contacts, $V_{12}$, $V_{34}$, on the sides  and $V_{56}$ on the centerline of the sample with silver-pasted current contacts on the other sides to test the inhomogeneous spatial distribution of the current.
We have observed the negative LMR ($I \parallel B$) and positive TMR ($I\perp B$) for all contacts V$_{12}$, $V_{34}$, and $V_{56}$,  as seen in Supplementary Fig.~S\ref{contact}. As shown there, we find that there is no significant variation of the voltage across the crystal.
Moreover, we have measured several samples with different sizes and thicknesses ($90\sim200$ $\mu$m). All samples with different size/thickness display comparable magnitude of the negative LMR as well as positive TMR (with $A$ remaining of the order of unity in all samples). Thus, the current jetting effect can be ruled out as the source for the magnetoresistance properties of Mn$_3$Sn.  

In conventional dirty semimetals and semiconductors, weak localization effects can cause negative magnetoresistance. Weak localization describes decrease in conductivity as the effect coming from constructive interference between two electron waves that travel along opposite directions along a closed path and are scattered off by the same impurities. Since an external magnetic field causes a phase difference between two waves, it disrupts the constructive interference, leading to enhanced conductivity or negative magnetoresistance~\cite{Hikami1980}. Just like any localization related effect, negative magnetoresistance due to field induced suppression of weak localization is generically a low temperature effect. In addition, for three dimensional materials, it has been demonstrated by Kawabata that both LMR and TMR become negative in the weak localization regime~\cite{Kawabata1980}. Thus weak localization cannot explain concomitantly observed negative LMR and positive TMR over a wide range of temperatures (50 K to 300 K), leaving the chiral anomaly of Weyl fermions as the bonafide mechanism behind our experimental observation of anisotropic magnetoresistance properties.  
\\

\subsection{Supplementary Note 8 : Electron doping effect due to extra Mn in off-stoichiometric Mn$_3$Sn}
\label{doping}
In the main text, we consider different levels of electron doping due to the extra Mn for ARPES and transport measurements, which are estimated by the theory as we describe in Methods.
To confirm the electron doping effect, we use ARPES to compare the spectra for Mn$_3$Sn single crystals with different Mn compositions.
In Supplementary Fig.~S\ref{arpes_doping}, we summarize the APRES results for Mn$_{3.02}$Sn$_{0.98}$ and Mn$_{3.10}$Sn$_{0.90}$.
By the theoretical calculation for the electron doping effect (Methods), the shift of $E_{\rm{F}}$ in Mn$_{3.02}$Sn$_{0.98}$ and Mn$_{3.10}$Sn$_{0.90}$ samples is estimated to be 12~meV and 59~meV, respectively.
The data show that the overall spectrum for Mn$_{3.10}$Sn$_{0.90}$ slightly shifts to a lower energy with respect to that for Mn$_{3.02}$Sn$_{0.98}$.
The shift of $E_{\rm{F}}$ is found to be $\sim$ 10~meV, which is quantitatively consistent with the theoretical estimate, 9.4~meV under the band renormalization of 5.
These results indicate that the extra Mn dopes electrons and the theoretical estimate works well at least in these doping range. 
%

%

\newpage
\begin{suppfigure}
	\begin{center}
		\includegraphics[width=0.8\columnwidth]{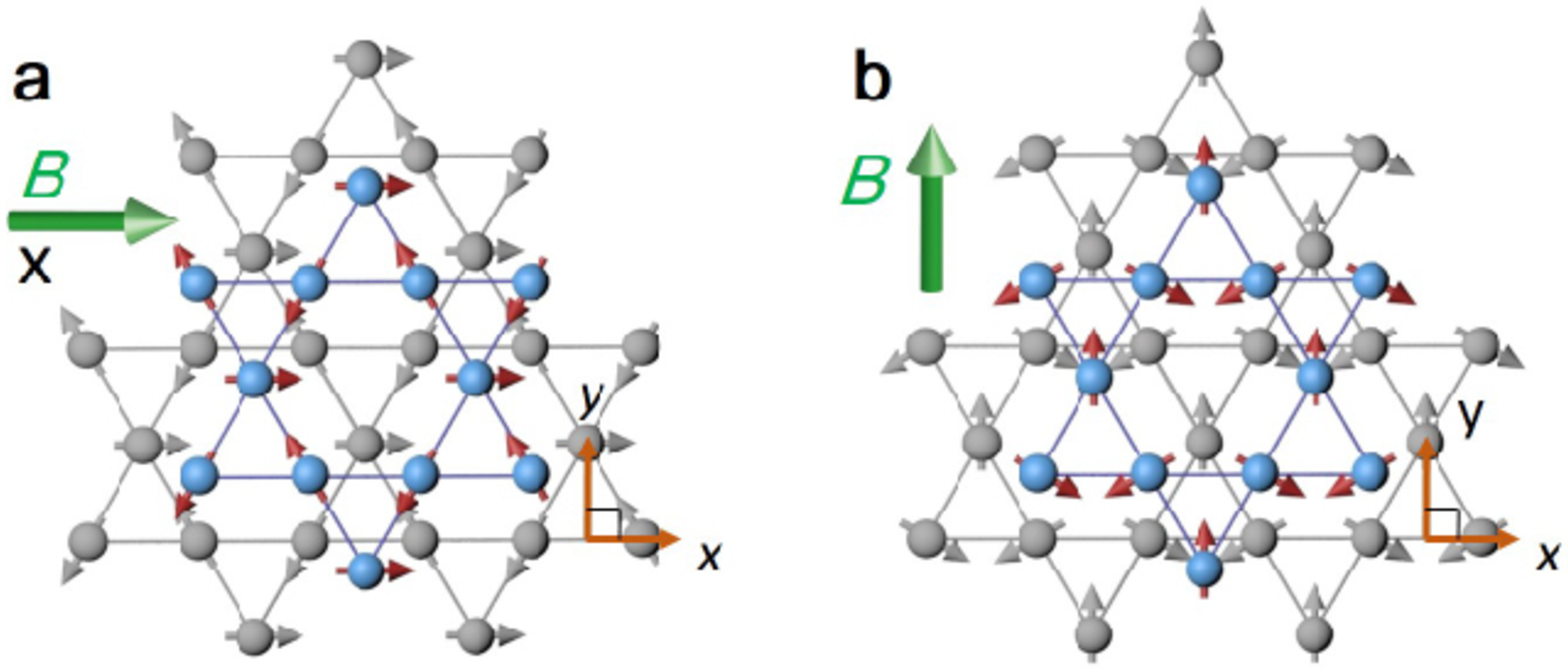}
		\caption{
			{\bf Antiferromagnetic structures in Mn$_3$Sn.}
			120~$^{\circ}$ magnetic structures of Mn local moments on the kagome lattice formed by Mn atoms (sphere).
			The neighboring two layers with $z$=0 (gray), $z$=1/2 (color) are shown. {\bf{a-b}} Magnetic structures in the field along {\bf{a}}, [$2\bar{1}\bar{1}$0] and {\bf{b}}, [01$\bar{1}$0]~\cite{Tomiyoshi1982}.
			Here, we define the $x$, $y$, and $z$ axes as [$2\bar{1}\bar{1}$0], [01$\bar{1}$0], and [0001], respectively.
		}
		\label{texture}
	\end{center}
\end{suppfigure}

\newpage
\begin{suppfigure}
	\begin{center}
	\includegraphics[width=0.7\columnwidth]{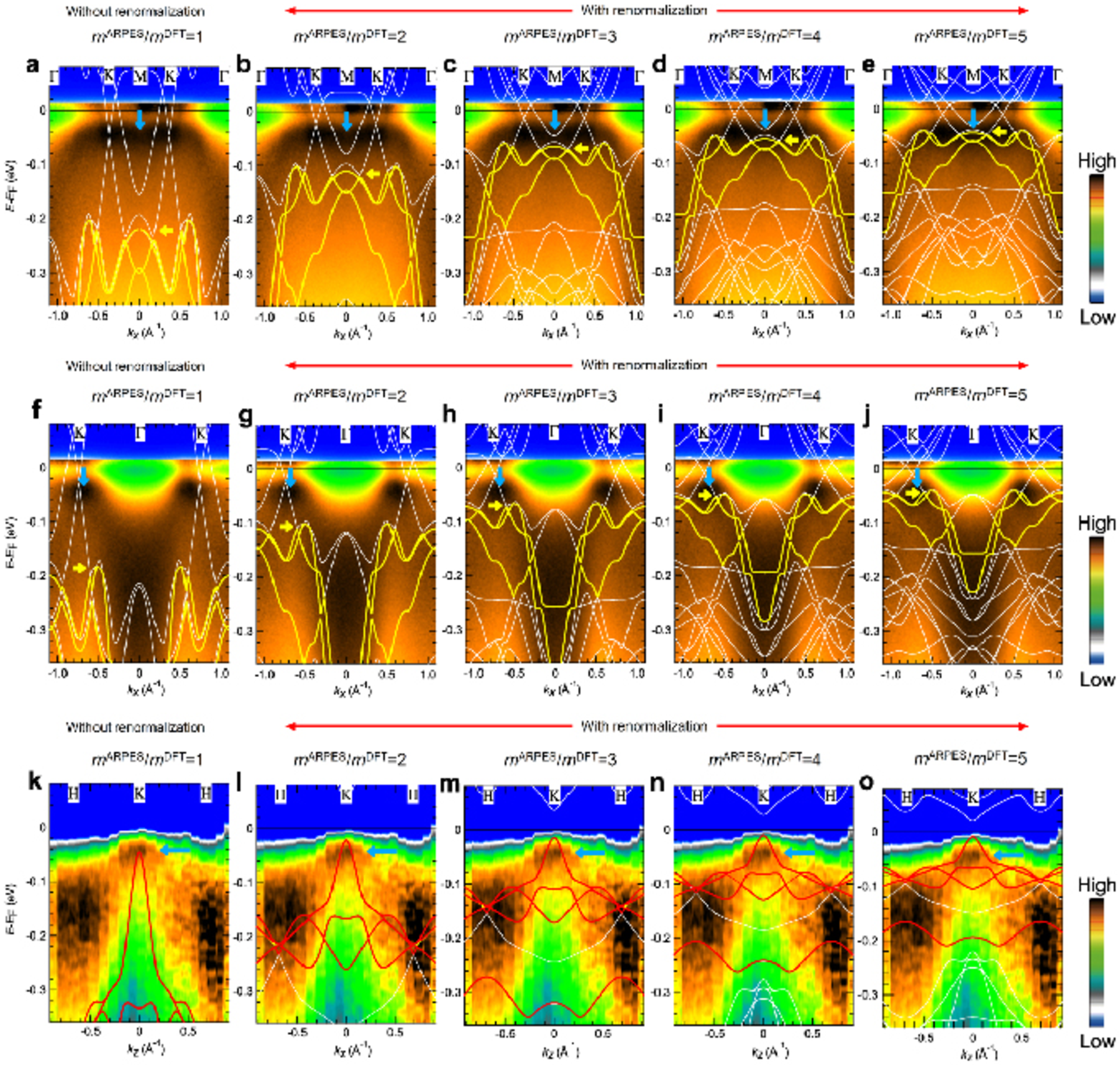}
	\caption{
		{\bf Strong electron correlation effects observed by ARPES.}
		ARPES band maps along ({\bf{a-e}}) $\rm{K}$-$\rm{M}$-$\rm{K}$, ({\bf{f-j}}) $\rm{K}$-$\rm{\Gamma}$-$\rm{K}$ and ({\bf{k-o}}) $\rm{H}$-$\rm{K}$-$\rm{H}$, which are compared with theoretical band dispersions renormalized by various band renormalized factors ($m^{\rm ARPES}/m^{\rm DFT}$) ranging from 1 to 5.
		The ARPES intensities shown in ({\bf{a-j}}) are divided by Fermi-Dirac distribution functions, and we find clear quasiparticle peaks near $E_{\rm{F}}$.  
		The energy position of the peak is marked by blue arrows. For $k_z$ band mapping in ({\bf{k-o}}), the ARPES intensities are shown without the Fermi-Dirac division to avoid erroneous analysis due to the variation of energy resolutions for different photon energies.  
		We find that the theoretical bands shown by the yellow bold lines and yellow arrows match the ARPES signals in both ({\bf{a-e}}) and ({\bf{f-j}}) when $m^{\rm ARPES}/m^{\rm DFT}$=5.
		The theoretical dispersions (red bold lines) renormalized by this value also explain the observed $k_z$ dispersion of the peaks shown in ({\bf{k-o}}).
	}
	\label{Ekmap_calc}
	\end{center}
\end{suppfigure}

\newpage
\begin{suppfigure}
	\begin{center}
		\includegraphics[width=0.90\columnwidth]{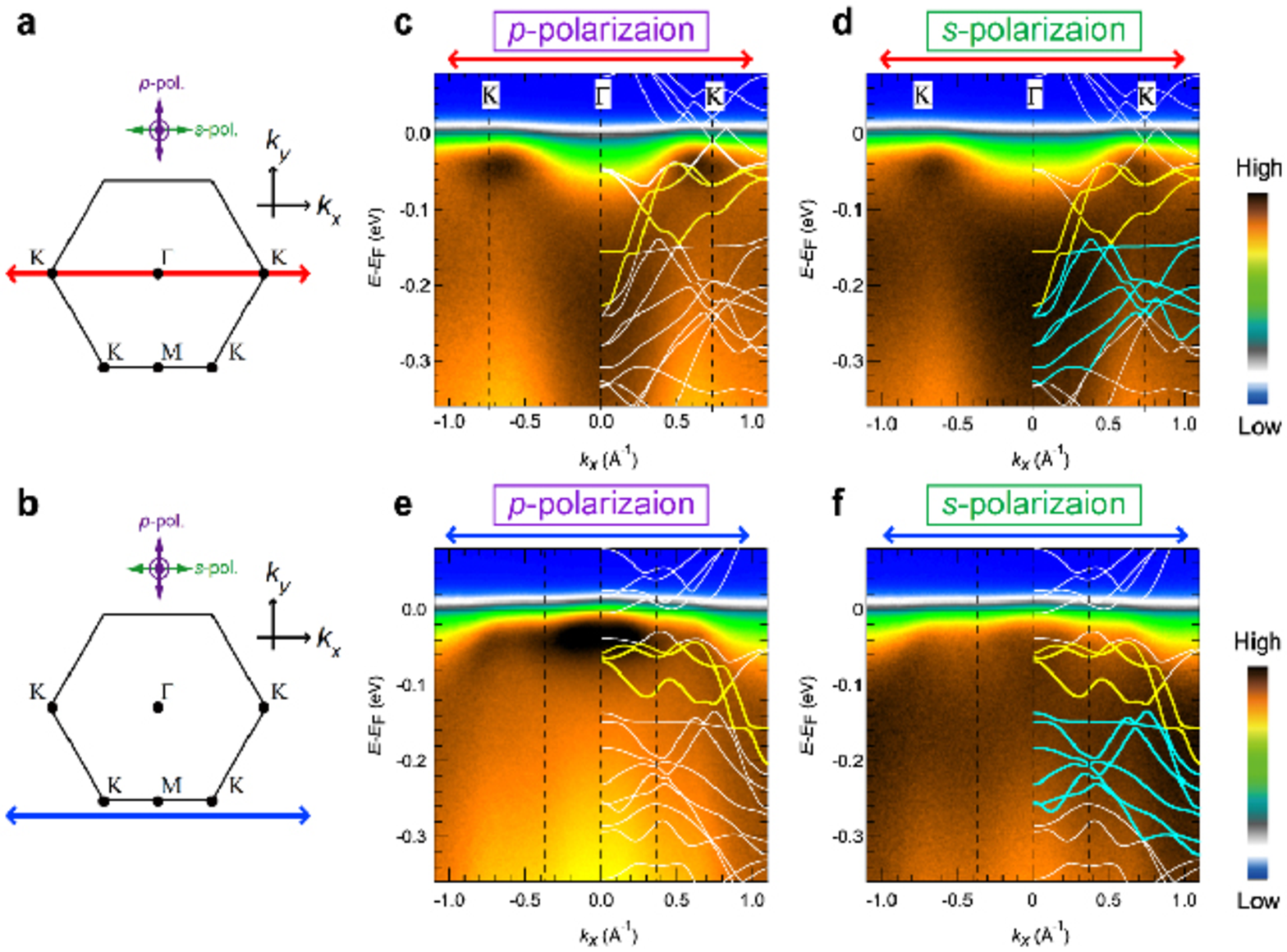}
		\caption{
			{\bf $p$- and $s$-polarization dependence of ARPES intensity maps.} 
			{\bf{a, b}}, Momentum sheet at $h\nu$=103~eV cut corresponds to the $k_x$-$k_y$ plane at $k_z$=0 and momentum cuts (red and blue arrows) along which ARPES data were obtained.
			The electric fields of $p$- and $s$-polarized incidence light are shown in the inset. 
			{\bf{c, d}}, ARPES band maps along K-$\Gamma$-K observed by $p$-polarized and $s$-polarized light, respectively.
			{\bf{e, f}}, ARPES band maps along $k_x$ slightly off from the high symmetry K-M-K line obtained with different light polarizations.
			These ARPES maps are compared with the theoretical band dispersions renormalized by a factor of 5.
			The yellow and sky-blue bold lines indicate the specific theoretical bands that likely dominate spectral intensities.
		}
		\label{polarization}
	\end{center}
\end{suppfigure}

\newpage
\begin{suppfigure}
	\begin{center}
		\includegraphics[width=0.98\columnwidth]{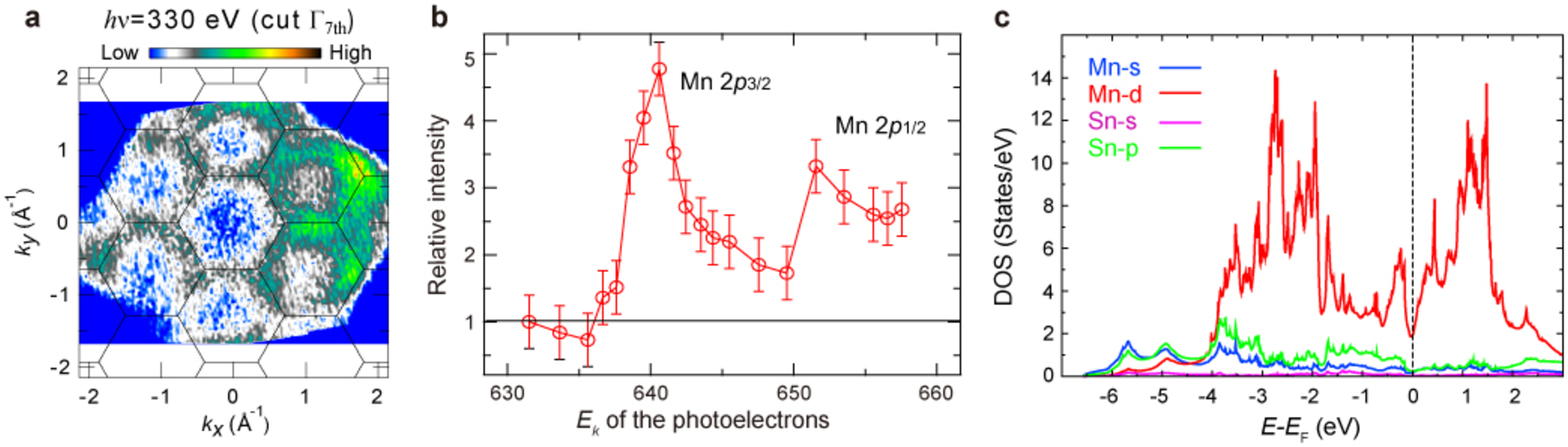}
		\caption{
			{\bf Mn 3$d$ bands obtained by soft X-ray ARPES (SX-ARPES).} 
			{\bf{a}}, Photoelectron mapping at a wide $k_x$-$k_{y}$ plane at $k_z$=0 obtained by SX-ARPES with $h\nu$=330~eV to cut the same $k_x$-$k_{y}$ plane as observed by $h\nu$=103~eV crossing $\rm{\Gamma}$.
			{\bf{b}}, Variation of valence photoemission intensity integrated within 50~meV below $E_{\rm{F}}$ as a function of $h\nu$.
			The intensity was normalized with that for $h\nu$=632~eV.
			The resonance at 640~eV and 652~eV corresponds to the transition from Mn 2$p$ to Mn 3$d$.
			{\bf{c}}, Total and partial density of states with SOC.
			The valence and conduction bands of Mn$_3$Sn are mainly composed of Mn 3$d$ states with small Mn 4$s$, Sn 4$s$ and Sn 4$p$ characters.
		}
		\label{resonance}
	\end{center}
\end{suppfigure}

\newpage
\begin{suppfigure}
	\begin{center}
		\includegraphics[width=0.9\columnwidth]{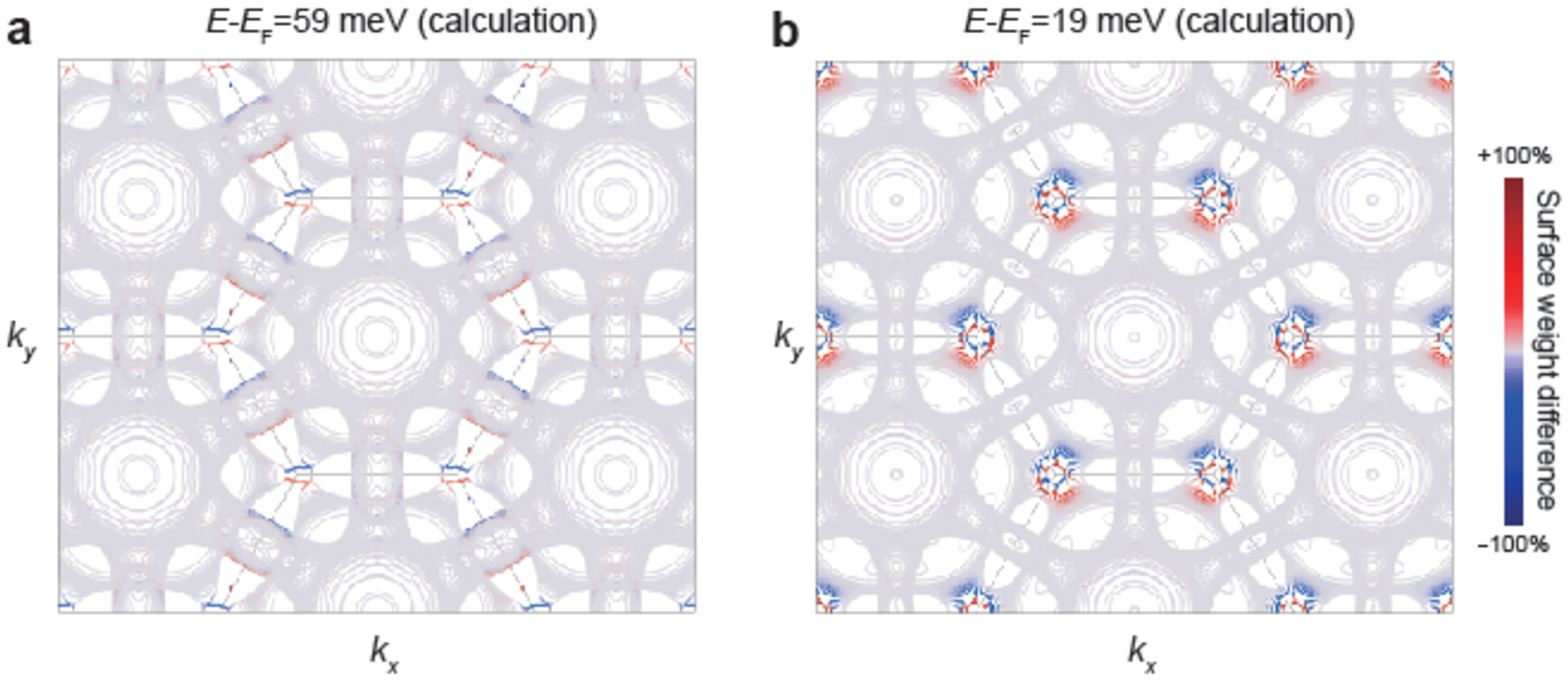}
		\caption{
			{\bf Calculated Fermi-arc surface states.}
			Surface spectral function of the Mn$_3$Sn~(0001) surface for the magnetic structure with the easy axis along $x$ (Supplementary Fig.~S\ref{texture}a) at {\bf a} 59~meV and {\bf b} 19~meV above $E_{\rm{F}}$ in the theoretical energy.
			The latter energy corresponds to the position of the shifted $E_{\rm{F}}$ estimated for Mn$_{3.03}$Sn$_{0.97}$ used in ARPES measurements.
			The surface Brillouin zone is shown by black lines.
			The Fermi arcs are shown by color scale, which are connected to the magnetic Weyl nodes.
			Red and blue indicate the surface state on the top
			and bottom surfaces, respectively.
			The gray area indicates the bulk projections.
		}
		\label{arc}
	\end{center}
\end{suppfigure}


\newpage
\begin{suppfigure}
		\begin{center}
			
	\includegraphics[width=0.5\columnwidth]{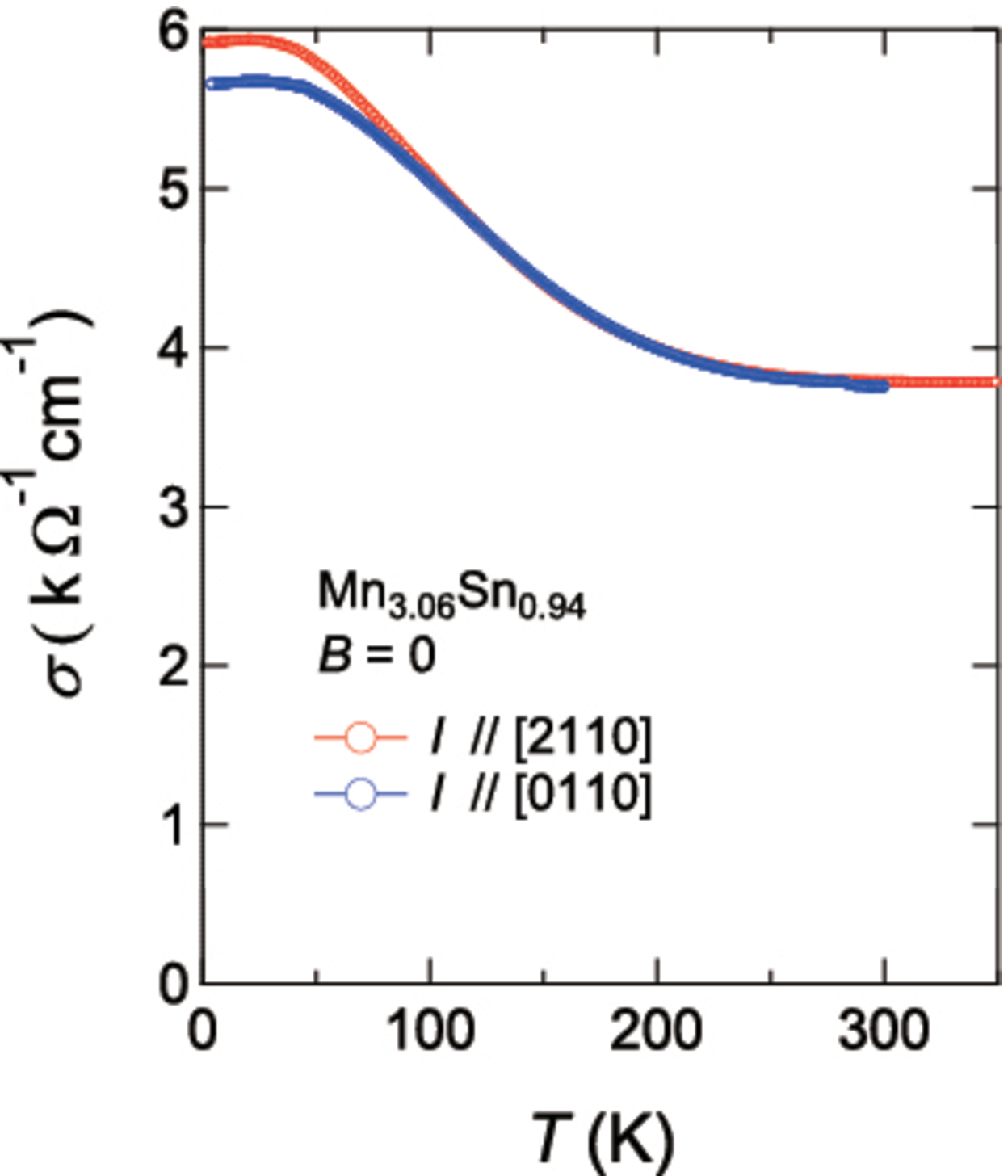}
	\caption{
		{\bf Longitudinal conductivity in Mn$_3$Sn.}
		{Temperature dependence of the longitudinal conductivity $\sigma = 1/\rho$ measured at 0 T with electrical current $I$ along [2$\bar{1}\bar{1}$0] (red) and [01$\bar{1}$0] (blue).}
	}
	\label{Long}
\end{center}
\end{suppfigure}

\newpage
\begin{suppfigure}
		\begin{center}
	\includegraphics[width=0.45\columnwidth]{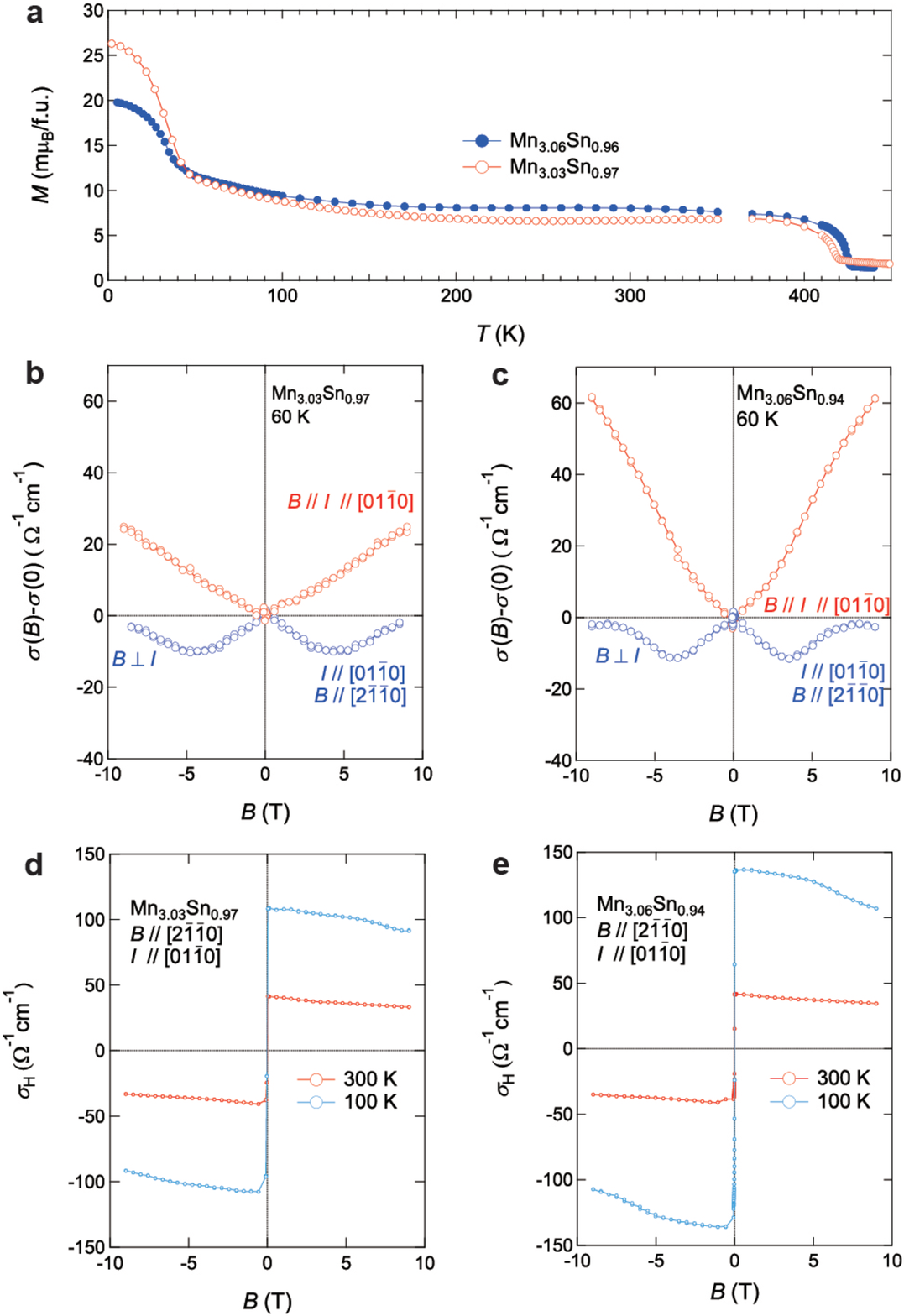}
	\caption{
		{\bf Magnetization, Longitudinal and Hall conductivity in Mn$_3$Sn.}
		{\bf{a}}, Temperature dependence of the magnetization measured in a magnetic field of 0.1 T along [2$\bar{1}\bar{1}$0] for Mn$_{3.03}$Sn$_{0.97}$ and Mn$_{3.06}$Sn$_{0.94}$ under a field-cooling sequence. The weak ferromagnetic components appear below the antiferromagnetic transition temperatures of 420~K and 426~K for Mn$_{3.03}$Sn$_{0.97}$ and Mn$_{3.06}$Sn$_{0.94}$, respectively. The anomaly around 50~K is due to the transition into the low-temperature cluster glass phase~\cite{Tomiyoshi1986, Feng2006}.
		{\bf{b-c}}, Magnetic field dependence of the magnetoconductivity $\sigma(B)-\sigma(0)$ at 60 K for $I \parallel B$ (red) and $I \perp B$ (blue) for {\bf{b}}, Mn$_{3.03}$Sn$_{0.97}$ and for {\bf{c}}, Mn$_{3.06}$Sn$_{0.94}$. 
		{\bf{d-e}}, Field dependence of the Hall conductivity $\sigma_{\rm H}$ measured at 300 K (red) and 100 K (blue) in magnetic field $B$ along  [2$\bar{1}\bar{1}$0] with electrical current $I$ along [0$1\bar{1}$0] for {\bf{d}}, Mn$_{3.03}$Sn$_{0.97}$ and for {\bf{e}}, Mn$_{3.06}$Sn$_{0.94}$. 
	}

	\label{Hall}
\end{center}
\end{suppfigure}

\newpage
\begin{suppfigure}
	\begin{center}
	\includegraphics[width=0.45\columnwidth]{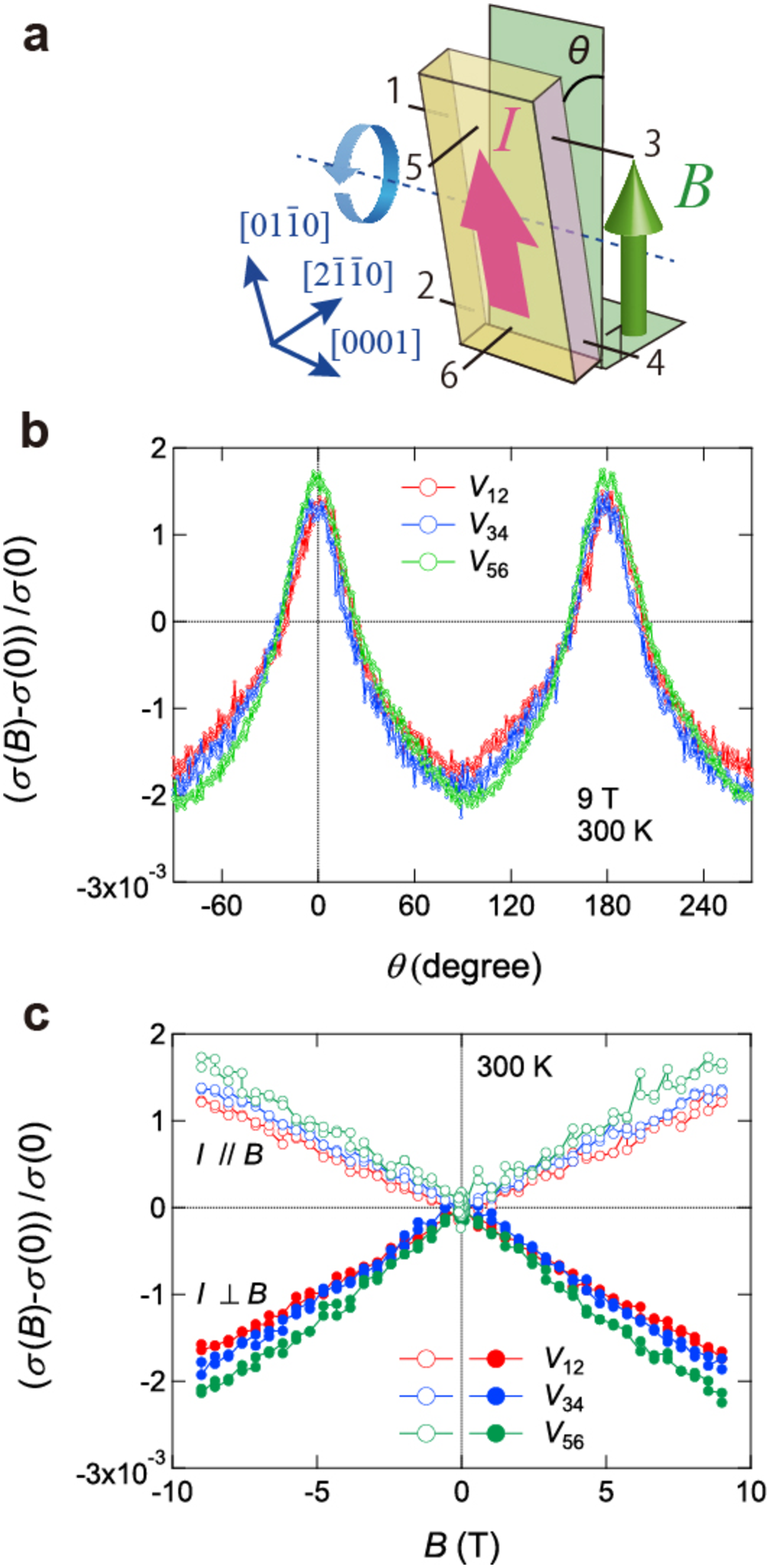}
	\caption{
		{\bf Longitudinal magnetoconductance measured using different voltage terminals.}
		{\bf{a}}, Schematic picture to illustrate the voltage terminal positions on a single crystalline sample of Mn$_3$Sn. 
		{\bf{b}}, Magnetoconductivity $(\sigma(B)-\sigma(0))/\sigma(0)$ measured under a field of 9 T as a function of angle $\theta$
		using different sets of the terminals 1-2 ($V_{12}$), 3-4 ($V_{34}$) and 5-6 ($V_{56}$) shown in the panel {\bf{a}}.
		{\bf{c}}, Field dependence of $(\sigma(B)-\sigma(0))/\sigma(0)$ measured for both $I \parallel B$ and $I \perp B$ at 300 K using the terminals 1-2 (side A), 3-4 (side B), and 5-6 (centerline)}.
	\label{contact}
\end{center}
\end{suppfigure}

\newpage
\begin{suppfigure}
	\begin{center}
		\includegraphics[width=0.95\columnwidth]{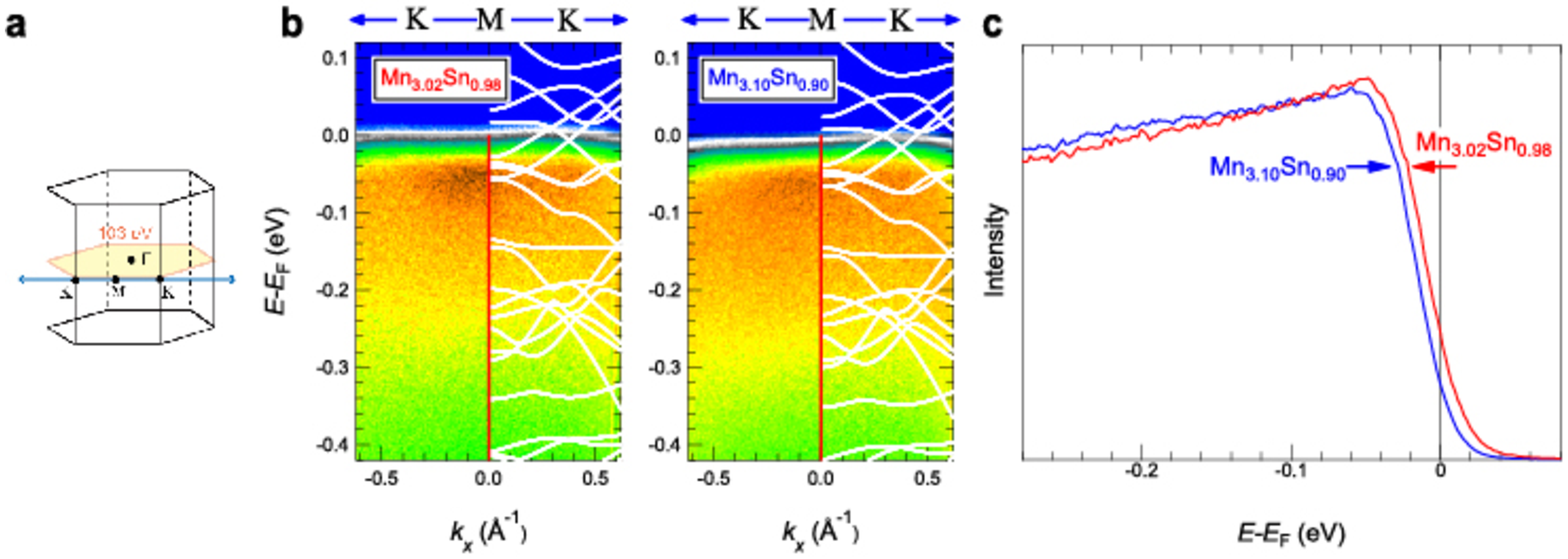}
		\caption{
			{\bf ARPES results for Mn$_3$Sn with different levels of electron doping.}
			{\bf a}, Brillouin zone, showing a momentum cut along which ARPES data were obtained. The momentum cut with $h\nu$=103~eV corresponds to the high symmetry line along {\rm{K}}-{\rm{M}}-{\rm{K}} (blue line).
			{\bf{b}}, ARPES band maps along the high-symmetry lines as shown in {\bf a}, for Mn$_3$Sn single crystals with different Mn doping, (left) Mn$_{3.02}$Sn$_{0.98}$ and (right) Mn$_{3.10}$Sn$_{0.90}$.
			According to theory, the extra Mn doping moves up $E_{\rm{F}}$ by 12~meV and 59~meV for Mn$_{3.02}$Sn$_{0.98}$ and  Mn$_{3.10}$Sn$_{0.90}$, respectively.
			These APRES intensity maps are compared with the calculated bands after shifting $E_{\rm{F}}$ and renormalizing them by a factor of $\sim$5.
			{\bf c}, Comparison of energy distribution curves obtained at M point as indicated by the energy dispersion in {\bf b}.
		}
		\label{arpes_doping}
	\end{center}
\end{suppfigure}

\end{document}